\documentclass[10pt,sigconf]{acmart}

\usepackage{graphicx}
\usepackage{color}
\usepackage{subcaption}
\usepackage{multicol}
\usepackage{amsmath,amsfonts,amsthm}
\usepackage{dsfont}
\usepackage{bbm}
\usepackage{algorithm}
\usepackage{algpseudocode}
\usepackage{xurl}

\usepackage{cleveref}
\usepackage{colortbl}

\newcommand{\DoeVAshcroft}{\emph{Doe v.\ Ashcroft}}
\newcommand{\DoeVGonzales}{\emph{Doe v.\ Gonzales}}

\renewcommand{\authors}{Alex Bellon, Miro Haller, Andrey Labunets, Enze Liu, Stefan Savage}


\begin{document}

\title[An Empirical Analysis on the Use and Reporting of NSLs]{An Empirical Analysis on the Use and Reporting of National Security Letters}

\author{Alex Bellon}
\email{abellon@ucsd.edu}
\affiliation{%
\institution{UC San Diego}
\city{La Jolla, CA}
\country{USA}
}

\author{Miro Haller}
\email{mhaller@ucsd.edu}
\affiliation{%
\institution{UC San Diego}
\city{La Jolla, CA}
\country{USA}
}

\author{Andrey Labunets}
\email{alabunets@ucsd.edu}
\affiliation{%
\institution{UC San Diego}
\city{La Jolla, CA}
\country{USA}
}

\author{Enze Liu}
\email{e7liu@ucsd.edu}
\affiliation{%
\institution{UC San Diego}
\city{La Jolla, CA}
\country{USA}
}

\author{Stefan Savage}
\authornote{Authors listed alphabetically.}
\email{ssavage@ucsd.edu}
\affiliation{%
\institution{UC San Diego}
\city{La Jolla, CA}
\country{USA}
}

\copyrightyear{2025}
\acmYear{2025}
\setcopyright{cc}
\setcctype{by}
\acmConference[CSLAW '25]{Symposium on Computer Science and Law}{March 25--27, 2025}{M{\"u}nchen, Germany}
\acmBooktitle{Symposium on Computer Science and Law (CSLAW '25), March 25--27, 2025, M{\"u}nchen, Germany}
\acmDOI{10.1145/3709025.3712209}
\acmISBN{979-8-4007-1421-4/25/03}

\begin{abstract}

Government investigatory and surveillance powers are important tools for examining crime and protecting public safety.
However, since these tools must be employed in secret, it can be challenging to identify abuses or changes in use that could be of significant public interest.
In this paper, we evaluate this phenomenon in the context of National Security Letters (NSLs).
NSLs are a form of legal process that empowers parts of the United States federal government to request certain pieces of information for national security purposes.
After initial concerns about the lack of public oversight, Congress worked to increase transparency by mandating government agencies to publish aggregated statistics on the NSL usage and by allowing the private sector to report information on NSLs in transparency reports.  
The implicit goal is that these transparency mechanisms should deter large-scale abuse by making it visible.
We evaluate how well these mechanisms work by carefully analyzing the full range of publicly available data related to NSL use.
Our findings suggest that they may not lead to the desired public scrutiny as we find published information requires significant manual effort to collect and parse data due to the lack of structure and context.
Moreover, we discovered mistakes (subsequently fixed after our reporting to the ODNI), which suggests a lack of active auditing. 
Taken together, our case study of NSLs provides insights and suggestions for the successful construction of transparency mechanisms that enable effective public auditing.

\end{abstract}

\maketitle

\renewcommand{\shortauthors}{A.~Bellon, M.~Haller, A.~Labunets, E.~Liu, S.~Savage}
\section{Introduction}\label{sec:introduction}
Intelligence and surveillance statutes such as the Foreign Intelligence Surveillance Act of 1978 and the USA PATRIOT Act of 2001 provision powerful legal tools that enable the government to collect a broad array of information for investigating crimes and protecting public safety. 
Due to the sensitive nature of such investigations, information about the content, target, or even existence of such requests may be security critical.
However, without any public information on their use, there is no oversight to ensure proper use and deter abuse.
Indeed, incidents of abuse have been documented in the past, such as the NSA's bulk collection of customers' telecommunication records~\cite{guardian-snowden}. 
One approach to address these concerns is increased transparency---either via statutory requirements for disclosure or documentation (e.g., mandated disclosure of wiretaps or aggregated data about the use of NSLs) or via voluntary reporting by the private sector (e.g., in transparency reports). 
The implicit reasoning is that even without revealing the specifics of individual operations, large-scale abuses will become evident under careful public scrutiny, and will thus deter such behavior.

However, the effectiveness of this approach hinges upon two assumptions: first, that such disclosures contain sufficient data, 
documentation and context that an unprivileged third-party could easily audit them for compliance, and second, that there are third-parties with the funding, focus and expertise necessary to do such work on a regular basis.
In this work, we investigate the extent to which these two assumptions hold in the context of National Security Letters~(NSLs).


We show that both assumptions may be too strong in practice.
Although both the public and private sector have sought to provide transparency by documenting NSL usage in aggregate, we document the need for better curation and documentation in this data, as evidenced by the significant amount of manual effort required to collect, parse, and process it. 
Additionally, we discovered data discrepancies that were subsequently fixed after reporting to ODNI, suggesting the lack of active auditing and checking from the public. 
We discussed with the former and current Chief, ODNI Civil Liberties, Privacy, and Transparency Office to validate our findings and understand current operational challenges as well as potential future improvements.

Altogether, our work evaluates existing transparency mechanisms for NSLs.
We provide empirical data by combining public resources and insights into operational challenges and real-world constraints that future work can consider when exploring new systems and approaches for accountability.
We suggest improvements to address some challenges without changing the balance between national security and transparency that the Congress decided on for NSLs.

 
Concretely, our paper makes the following contributions:
\begin{itemize}
    \item
    We present the first dataset that consolidates heterogeneous NSL data from government statistics, transparency reports, and published NSLs. 
    With a combination of manual effort and automated scripts, we produce a normalized dataset that enables comparisons across data sources. 
    We make this data available, both to enable future research and allow for easier auditing by the public.

    \item As a result of our data collection, we identify and report data inconsistencies present in the datasets to the relevant government agencies, which have since been corrected~\cite{fisa-correction-2020}.
    
    
    \item 
    We analyze this data and provide insights about the use and reporting of NSLs over time, including the following:
    NSL requests for non-US persons grew significantly between 2010 and 2015, passing the number of requests for US persons, and remaining popular until today, with significant spikes in frequency in 2015 and 2019.
    Furthermore, our analysis of public transparency reports suggest that telecommunication companies (e.g., AT\&T, T-Mobile, and Verizon) receive the largest number of NSLs among the reporting companies.
    Our cross-comparison of transparency reports with the number of NSL requests reported by the government appears to be consistent.
    
    \item
    We provide suggestions for improving the overall environment of NSL reporting in both public and private sectors. 
    We outline changes that can be made to the government data publishing process that can improve the accuracy and security of data without adding undue burden. 
    In addition, we offer suggestions for improved transparency in the private sector for NSL-related data publishing. 
\end{itemize}

\section{Related Work}
There are four main lines of research that are related to our work. 
First, a body of work argues for the importance of transparency in discouraging potential abuse of power by the government, such as judicial power~\cite{smith2009kudzu} and surveillance activities~\cite{smith2012gagged, bloch2018exposing}.
These discussions inspired us to empirically evaluate the transparency mechanisms of NSLs.

A second area of research suggests novel mechanisms~\cite{frankle2018practical, segal2014catching, krollsecure} that enable transparency and accountability while preserving secrecy. 
These mechanisms utilize cryptographic primitives and often allow auditing at a fine-grained (e.g., per-case) level without disclosing confidential information or violating privacy.
Instead of exploring new mechanisms with optimal transparency and accountability properties, our work evaluates existing mechanisms, operational practices, and challenges.
We hope that future work and discussions in Congress can combine our empirical understanding of deployed mechanisms with new ideas---such as~\cite{frankle2018practical, segal2014catching, krollsecure}---to codify new mechanisms that are practical and effective.

A third research direction related to our work encompasses the empirical analysis of publicly available data.
This includes Romanosky et al.'s~\cite{romanosky2014empirical} analysis of data breach litigations, Beller's~\cite{beller2022401} analysis of publicly available notices issued to targets being spied on under the authorization of FISA, and Kesari's~\cite{kesari2023data} assessment of identity theft report rates. 
Our work also takes an empirical approach and utilizes data to gain insights into the state of transparency of NSLs.

Finally, there are various discussions of NSLs in the legal research community.
However, most of them are centered around the constitutionality of NSL provisions in the context of First and Fourth Amendment issues~\cite{bloch2016process, dallal2018speak, garlinger2009privacy, gorham2008national, nieland2006national, schrop2017your}. 
One exception is the analysis by EPIC~\cite{EPICNati14:online}, which also collects and examines public data.
However, their analysis and data collection is limited in depth and scope, and focuses mostly on events between the PATRIOT Act in 2005 and the USA FREEDOM Act in 2015.

\section{NSL: A Brief Background}\label{sec:nsl_history}

Currently, five statutory provisions authorize government agencies to issue NSLs: the Right to Financial Privacy Act~\cite{law:USC-12-Chapter-35-RFPA}; the Electronic Communications Privacy Act~\cite{law:USC-18-2701-ECPA-access-stored-communications}; the National Security Act~\cite{law:PL-80-253-NSA}; the Fair Credit Reporting Act~\cite{law:USC-15-1681-FCRA}; and the USA PATRIOT Act~\cite{law:PL-107-56-PATRIOT} (the last of which being the first to codify the term ``National Security Letter'' explicitly).
Together, these statutes allow government agencies---predominantly the FBI---to use NSLs to request metadata for a person (often referred to as ``target'') from various companies including electronic communication service providers and credit agencies.
A single NSL can contain multiple Requests For Information (ROI)---such as an account identifier or email address---as long as they are in the context of a single investigation.
Additionally, the FBI may serve multiple NSLs for the same person under different statutes and to different entities (e.g., to collect information on phone calls, emails, and financial records).
The reported usage statistics distinguish NSLs for US persons, non-US persons, and subscriber information. 
The latter includes NSL letters for both US and non-US persons where the target's nationality was not known at the time when the NSL letter was issued.

In this section, we start by giving a short overview of the five NSL statutes, followed
by a summary of efforts to improve transparency around the use of NSLs (notably two
amendments to the five NSL statutes that introduced reporting mandates). We refer the
interested reader to \Cref{sec:full_background} for a more extensive historic background on NSLs.

\subsection{The Five NSL Statutes}
The Right to Financial Privacy Act (RFPA)~\cite{law:USC-12-Chapter-35-RFPA} was the first statute used to introduce NSL authorities.  
As part of a 1986 amendment, the FBI was granted the right to request access to business records from financial institutions (amended in 2003 to include a  broader range of organizations).  
During the same period of time, the second NSL-granting statute, the Electronic Communications Privacy Act (ECPA)~\cite{law:USC-18-2701-ECPA-access-stored-communications} was enacted.
It provided access (via 18 U.S.C~\S~2709) to business records (i.e., name, address, length of service, and toll records) of wire or electronic communication service providers for counterintelligence purposes. 

In the 1990s, two more statutes were enacted granting further NSL authorities.
In 1994, the National Security Act (NSA) was amended to include a procedure for any authorized agency to request a broad array of business records from various organizations for investigating potential document leaks from government employees (codified at 50 U.S.C~\S~3162~\cite{law:USC-50-3162-codified-IAA}).
Later, in 1996, the Fair Credit Reporting Act (FCRA) was amended to incorporate the fourth statutory provision for NSLs, which authorized the FBI to access credit agency records in service of national security investigations.

Finally, the PATRIOT Act~\cite{law:PL-107-56-PATRIOT} is the fifth and last NSL-granting statute, which was enacted in 2001 as a response to the terrorist attacks of September 11~\cite{wikipedia:patriot-sept-11}.
Through amending FCRA, the PATRIOT Act introduced a new procedure for government agencies to access consumer reports from credit report agencies for counterintelligence and counterterrorism investigations.
In addition, it made substantial amendments to three of the four existing NSL statutes (RFPA, ECPA, and FCRA), expanding the scope of NSLs and simplifying the administrative approval requirements.  


\subsection{Reporting and Nondisclosure Mandates of NSLs}
Initially, the five NSL statutes had no clear mechanism for judicial review, relaxed reporting requirements, and strict nondisclosure mandates\footnote{%
    As an example, the original ECPA and FCRA both prohibited the disclosure of an NSL request to any person.
}.
Given the broad range of data accessible via NSLs, this unsurprisingly sparked concerns both in the public and, in response to challenges, in the courts~\cite{CRS:legalBackground}. 
Over time, partially fueled by unfavorable judicial reactions (e.g., \DoeVGonzales~\cite{case:DoeVGonzales}) and various public events (e.g., the Snowden disclosures in 2013~\cite{guardian-snowden}), several efforts have been made to increase the transparency of NSL usage~\cite{CRS:legalBackground}. 
We highlight the two legislative efforts that are most relevant to this paper.

In 2006, Congress amended the NSL statutes with the USA PATRIOT Improvement and Reauthorization Act~\cite{law:PL-109-177-PATRIOT-2005} and the USA PATRIOT Act Reauthorization Amendments Act~\cite{law:PL-109-178-PATRIOT-2005}.
Notably, the USA PATRIOT Improvement and Reauthorization Act added two \emph{unclassified} reporting requirements. 
Section 119 of the Act initiated an audit by the Office of Inspector General of the Department of Justice (DOJ) to report NSL usage to Congress (``the OIG reports''). However, these reports were each one-off occurrences, and the OIG is not expected to publish any further reports unless there is another mandate from Congress.
Second, Section 118 requires the Attorney General of the DOJ to submit annual unclassified statistics about NSLs targeting \emph{US persons} to Congress~\cite{law:PL-109-177-PATRIOT-2005-section-128}.
These statistics are published as part of the Foreign Intelligence Surveillance Act reports, hence referred to as ``FISA reports''.

In 2015, the USA FREEDOM Act~\cite{law:PL-114-23-FREEDOM} mandated additional reporting, and expressly limited the use of NSLs for explicitly specified information.
Notably, it mandates that the Office of the Director of National Intelligence (ODNI) publishes a report that details both the number of NSLs issued, and the number of requests for information contained in those NSLs for the past year (in addition to the FISA reports)~\cite{law:PL-114-23-FREEDOM}.
This is published as part of the National Intelligence's Annual Statistical Transparency Report (ASTR). 

Finally, in addition to adding reporting mandates and limiting the scope of NSLs, the USA FREEDOM Act introduced three major changes that relaxed the nondisclosure requirements. First, it allowed companies to report the total number of NSLs received (and customers covered by those NSLs) quantized into bands of 250, 500 or 1000.\footnote{%
    Public records~\cite{CRS:legalBackground, law:HRep-114-109} suggested that this part of the USA Freedom Act was inspired by the pre-existing voluntary agreement reached in 2014 between several technology companies and the DOJ~\cite{dojagreement:2014}.
}
In practice, a range of companies (particularly in the technology and communications sectors) take advantage of this permission and publish a rough estimate of the number of NSLs they receive as part of their annual transparency reports. We refer to such reports collectively as ``transparency reports''. Moreover, the USA FREEDOM Act codified a procedure (18 U.S.C.~\S~3511) that allows companies to request judicial review of the nondisclosure orders (commonly known as ``reciprocal notice''). Lastly, it required the Attorney General to adopt procedures to review nondisclosure orders at appropriate intervals. If disclosure is no longer believed to lead to harms listed in 18 U.S.C.~\S~2709(c)\cite{law:USC-18-2709-ECPA-access-to-records}, then the FBI should terminate the nondisclosure order and notify the company. 
Some companies have since chosen to publish the content of such NSLs (i.e., after the termination of nondisclosure requirements) which can provide useful metadata for analyzing past NSL issuance. We refer to such publications collectively as ``company NSLs''.

 

In summary, three types of NSL reports are currently available to the public: 
\begin{itemize}
    \item OIG, FISA and ASTR reports (collectively referred to as ``Government Reports'' in Section~\ref{subsec:gov_stats}) mandated by the USA PATRIOT Improvement and Reauthorization Act and USA FREEDOM Act
    \item Company transparency reports that include the number of received NSLs
    \item NSLs disclosed by companies after the NSL's nondisclosure requirement is lifted
\end{itemize}
Understanding what the public can learn from these publicly available NSL reports, how useful the information is to the public, and the limitations of each data source is the central focus of our work.

\section{Data Collection and Trends}\label{sec:data_collection}

\begin{figure*}
    \centering
    \includegraphics[width=\linewidth]{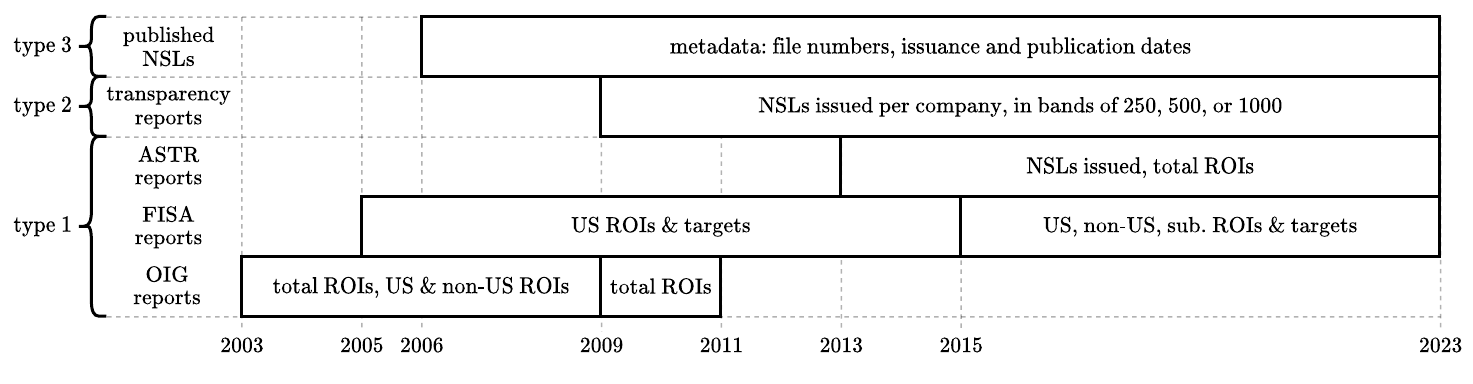}
    \caption{Overview of three types of data sources: government statistics, transparency reports, and published NSLs}\label{fig:nsl_data_srcs}
    \Description[Overview of three types of data sources: government statistics, transparency reports, and published NSLs.]{
        Overview of three types of data sources: government statistics, transparency reports, and published NSLs. 
        Type 1 contains ASTR, FISA, and OIG reports from the government. 
        ASTR reports publish the number of NSLs issues as well as the number of total ROIs since 2013.
        FISA reports include the number of ROIs and targets for US persons since 2005 and additionally publish information on non-US persons and subscriber only information requests starting from 2015.
        OIG reports published the total number of ROIs, as well as number of US and non-US ROIs between 2003 and 2009, and only the total number of ROIs between 2009 and 2011.
        Type 2 are transparency reports where companies report the number of NSLs they received in bands of 250, 500, or 1000.
        Type 3 are published NSLs containing metadata such as file numbers, as well as issuance and publication dates.}
\end{figure*}

Despite that the published data should improve transparency and auditability of NSL usage, interpreting and auditing these reports is a tiresome process.
Significant manual effort is required to identify, collect, clean, normalize and interpret the range of public NSL reports.
There exists no single collection of NSL information, and finding reports from various entities is not always straightforward or efficient. 
Additionally, there is no universal format that all reports follow. 
Company transparency reports are even less structured, as they are reported voluntarily and without any standards defining format, frequency of publication, or required information.  
Moreover, data may overlap, such as when NSLs with the same number are served to multiple companies.\footnote{%
    For instance, NSL-19-483160 is served to both Google and Apple.
}

One contribution of this paper is the first comprehensive collection of NSL statistics, transparency reports, and published NSL letters scattered across the Internet. 
Additionally, using automated scripts and manual cleaning, we extract all useful information from the NSL reports into a normalized and computer-friendly format, enabling several analyses detailed in later sections.
We make our data and scripts public.\footnote{%
    We publish our data set here:  \url{https://github.com/ucsdsysnet/nsl-empirical-analysis}.
}

\Cref{fig:nsl_data_srcs} shows the time ranges\footnote{%
    These ranges do not necessarily correspond to amendments.
    For instance, companies only received permission to publish statistics on the number of transparency reports they received after an agreement with the DOJ in 2014~\cite{dojagreement:2014}.
    However, after 2014, some companies retroactively published their NSL statistics back to 2009.
} and information that we were able to collect from different sources:
\begin{itemize}
    \item \emph{Type 1} (cf.~Section~\ref{subsec:gov_stats}):
    Multiple government entities publish statistics about NSL usage (\emph{type 1} data): 
    The OIG reports from 2006~\cite{ig-report-2006}, 2007~\cite{ig-report-2007}, and 2014~\cite{ig-report-2014} contain the sum of all types of ROIs for 2003--2011, and US and non-US ROIs for 2003--2009.
    FISA reports~\cite{fisa-reports} contain NSL requests and ROIs for US targets since 2005, and data for non-US and subscriber information requests since 2015.

    \item \emph{Type 2} (cf.~Section~\ref{subsec:transparency_reports}):
    Companies may choose to publish transparency reports but are restricted to reporting the number of NSL requests they receive in bands of 250, 500, or 1000.
    Although non-disclosure orders initially silenced companies, they were allowed to publish such reports in 2014, and some companies retroactively published reports back to 2009.

    \item \emph{Type 3} (cf.~Section~\ref{subsec:company_nsls}):
    After associated gag orders are lifted, companies may publish redacted NSL letters that they received.
    While personally identifiable information is redacted, the letters still contain useful metadata including file numbers and issuance dates.
    In most cases, companies document when they published an NSL.
\end{itemize}

Below, we detail how we collect and clean data from each source and observe general trends.

\subsection{Government Reports}\label{subsec:gov_stats}

\begin{figure*}[t]
    \centering
    \begin{subfigure}[b]{0.49\textwidth}
        \centering
        \includegraphics[width=\textwidth]{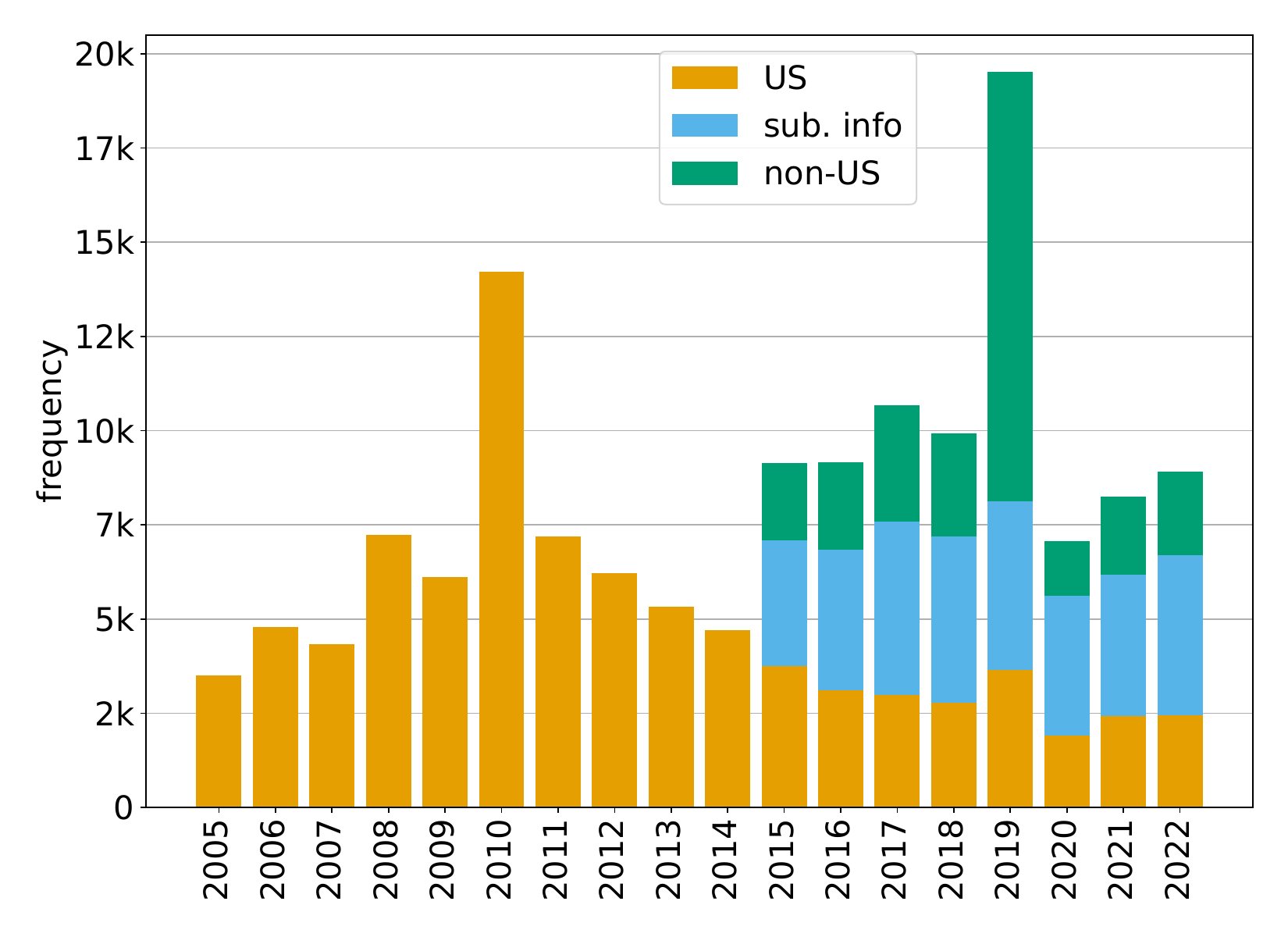}
        \caption{%
            Number of targets.
        }\label{fig:nsl_targets}
    \end{subfigure}
    \begin{subfigure}[b]{0.49\textwidth}
        \centering
        \includegraphics[width=\textwidth]{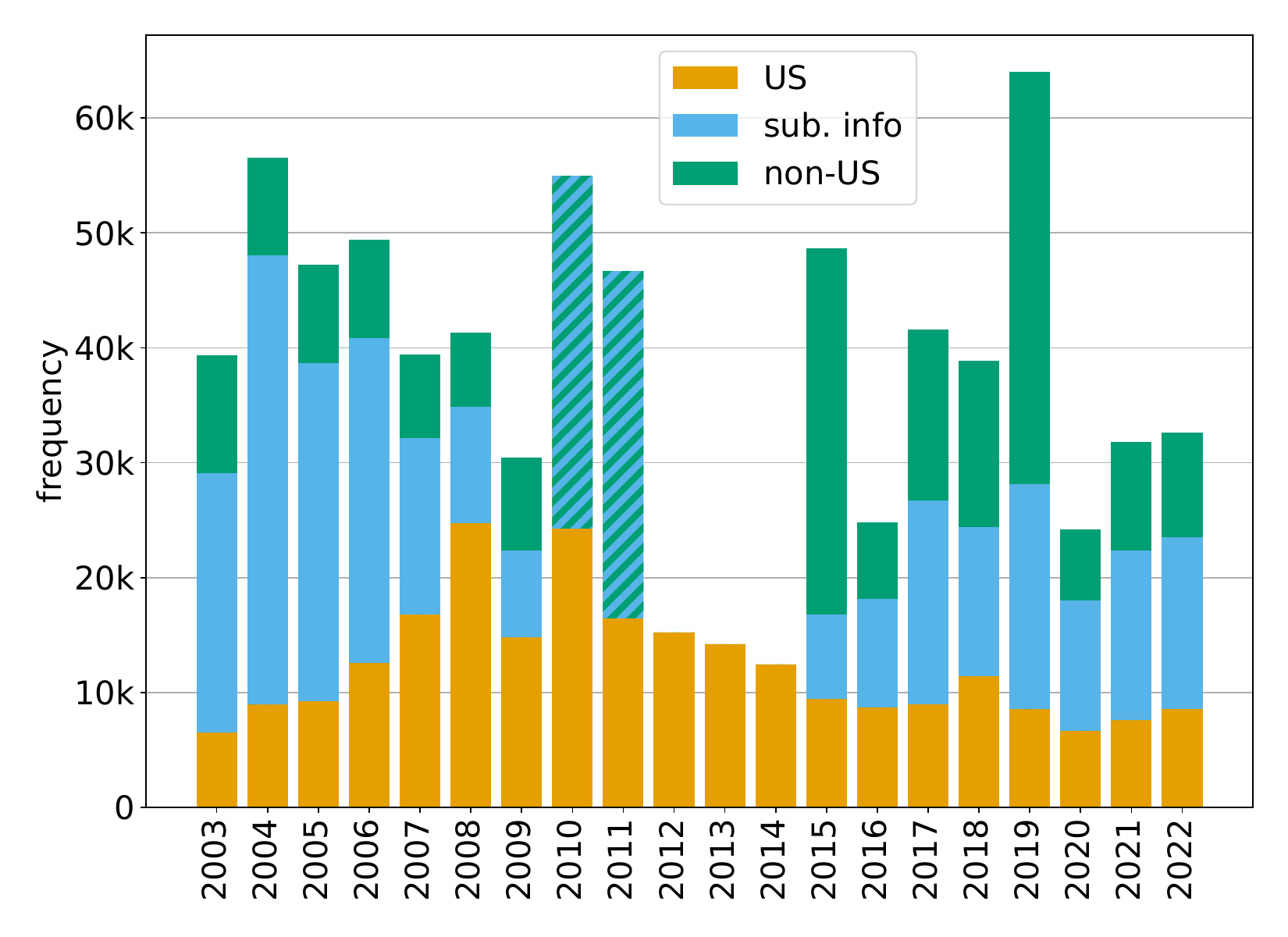}
        \caption{%
            Number of ROIs.
        }\label{fig:nsl_roi_requests}
    \end{subfigure}
    \caption{%
        Reported information about NSL requests for US persons, non-US persons, and subscriber information (abbreviated as ``sub. info'') in case the target's nationality is unknown at the time of request.
        A single NSL letter can request data for multiple ROIs (e.g., email addresses) of the same target.
    }
    \label{fig:nsl_requests}
    \Description[
        Two bar plots show the number of targets as well as the number of ROIs between 2003 and 2022.]{
        Reported information about NSL requests for US persons, non-US persons, and subscriber information (abbreviated as ``sub. info'') in case the target's nationality is unknown at the time of request.
        A single NSL letter can request data for multiple ROIs (e.g., email addresses) of the same target.
    }
\end{figure*}

NSL statistics collected from various government reports~\cite{astr,fisa-reports,ig-report-2006,ig-report-2007,ig-report-2014} include the number of NSL requests for US and non-US persons, as well as subscriber information for any person.
\Cref{fig:nsl_requests} depicts this data. 
The number of targeted individuals for every type of NSL request is enumerated in \Cref{fig:nsl_targets}.
\Cref{fig:nsl_roi_requests} shows the number of ROIs, which exceeds the number of targets as one person can have multiple NSLs issued under different statutes, each containing multiple ROIs.
While the published information does not allow us to associate NSL letters with their respective number of ROIs, we can give overall statistics.
Since 2015, there were on average 3.8 ROIs per target, with a standard deviation of~0.7. 
The maximum was in 2015, at 5.3 ROIs per target.

The FISA reports~\cite{fisa-reports,law:fisaReport} started to mandate reporting NSL requests for US persons in 2003. Since then, the format of the reported data has changed multiple times due to new regulations and policies.
In 2005, section 128 of the USA PATRIOT Improvement and Reauthorization Act added the requirement to report the number of ROIs for US persons made with NSLs~\cite{law:PL-109-177-PATRIOT-2005-section-128}.
Starting in 2015, the USA FREEDOM Act~\cite{law:PL-114-23-FREEDOM} additionally required the reporting of NSL requests for non-US persons and NSL requests for subscriber information.
The ASTR~\cite{astr} published by the ODNI reports the total number of issued NSLs and ROIs.

For 2003--2009, we found that reports of the OIG~\cite{ig-report-2007,ig-report-2014} contained statistics (that were redacted in an early version but disclosed in the 2014 revision~\cite{ig-report-2014}) from reports of the FBI to Congress under FISA on the number of ROIs in NSLs for US and non-US persons.
The OIG reports also include the total number of NSL requests until 2011, from which we can infer the number of subscriber information ROIs for 2003--2009 and the combined number of non-US ROIs and subscriber information requests for 2010 and 2011.
Unfortunately, the OIG reports do not contain any additional information about non-US or subscriber information NSLs.
Thus, for the period from 2012 to 2014, we only know statistics about ROIs pertaining to US persons due to the lack of declassified information (see \Cref{fig:nsl_roi_requests}).
Similarly, we do not have any information on the targets of NSLs for non-US and subscriber information before 2015 as shown in (see \Cref{fig:nsl_targets}).

\Cref{fig:nsl_requests} indicates that the number of NSL requests for US persons grew steadily until 2010, and then eventually decreased back to the level of 2003.
Partial data from OIG reports~\cite{ig-report-2007,ig-report-2014} suggests that NSL requests for non-US persons were in the minority before 2010, but became more common between 2010 and 2015 and still remain popular.
Additional data from the FISA reports shows that after 2015, the US-targeted NSL requests represent a smaller portion of all requests, although a significant portion of requests are made for subscriber information of any person.\footnote{It is tempting to relate fluctuations in NSL requests to global political events. For instance, the peak in 2019 for non-US ROIs may be due to increased investigations into digital attacks to perform economic espionage, which dominated the FBI's counterintelligence program in 2019 according to FBI director Christopher Wray~\cite{counterintelligence-2019}.  However, it is impossible to corroborate such a hypothesis using only the data available here.}

\subsection{Transparency Reports}\label{subsec:transparency_reports}

Transparency reports provide the perspectives of individual companies on NSL issuance. 
In the United States, there is neither a legal requirement nor a standardized format for publishing transparency reports.
Moreover, the government restricts the reporting of the number of NSL requests to bands of 250, 500, or 1000 as defined by 50~U.S.C.~\S~1874~\cite{law:USC-50-1874-NSL-ranges}. 
These restrictions and lack of structure cause the number, frequency, content, and accessibility of transparency reports to vary wildly by company and year.

We collect transparency reports published by 55 US-based private companies (listed in Appendix~\ref{subsec:data-src-transparency-reports}), of which 41 provided data related to NSLs.
These companies were selected from Access Now's Transparency Reporting Index\footnote{\url{www.accessnow.org/campaign/transparency-reporting-index}}, which to our knowledge is the only public database of companies that publish transparency reports. We removed 33 companies listed in the Index from consideration as they were not US-based companies, and therefore did not publish any information related to US national security inquiries. Given the lack of centralized data about transparency reports, our findings necessarily represent a lower bound for the number of NSL requests issued to US companies. 

While some of the companies have all of their transparency reports easily archived and downloadable as CSV or PDF files, other reports are hidden in blog posts or support forum answers, posted as low-quality images, or have broken links.
In addition, we had to exclude the reports of 17 companies as they only released aggregated numbers for all national security requests they received, mixing NSLs with other FISA requests (e.g., for electronic surveillance).
Our dataset is therefore not comprehensive and only provides lower bounds on NSL usage.


\begin{figure}[t]
     \centering
     \includegraphics[width=\linewidth]{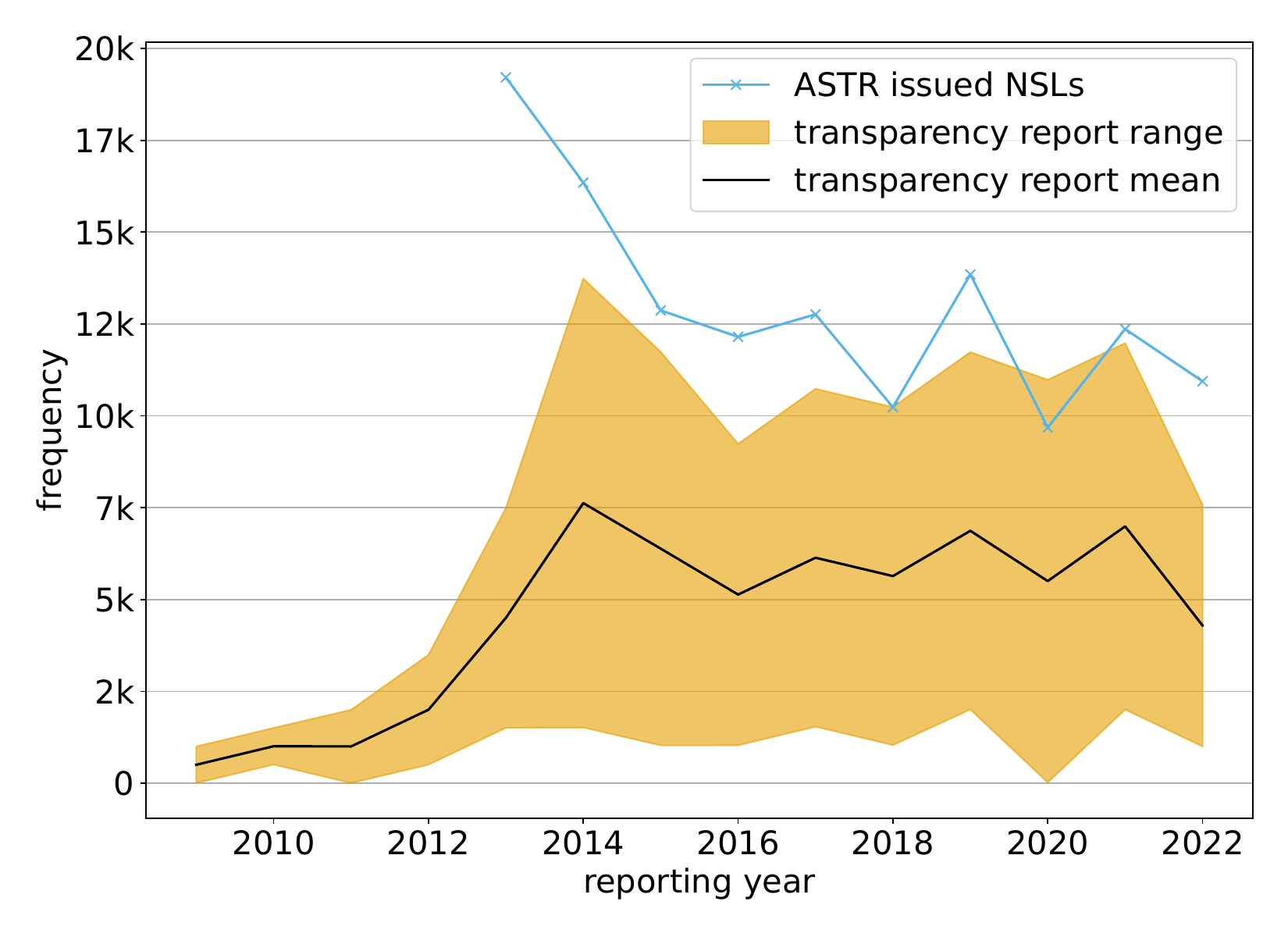}
     \caption{%
        The highlighted area shows the cumulative transparency report ranges, the sum of limits reported by some companies.
        We compare them to the number of issued NSLs reported by the ASTR.
    }
     \label{fig:compare-transparency-records-and-astr}
     \Description[This visualization shows the cumulative transparency report ranges, the sum of limits reported by some companies. We compare them to the number of issued NSLs reported by the ASTR.]{
        The plot shows that the number of issued NSLs is most of the times higher than the cumulative upper bound of the limits, showing that there is missing data.
        However, the inclines and declines seem consistent with the changes in the cumulative bounds.
     }
\end{figure}

Despite the incomplete data set and coarse reporting bands, we can still compare the respective sums of lower and upper bounds with the total number of issued NSLs from ASTR as shown in \Cref{fig:compare-transparency-records-and-astr}.
We find that the number of NSL requests from ASTR exceeds the cumulative upper bound from transparency reports, indicating that we are indeed missing a significant fraction of company reports for NSLs.
%
Although the company transparency records are not a comprehensive data source, we observe that after 2013, the mean of the bands strongly correlates with the contemporaneous number of issued NSLs (with a Pearson coefficient of 0.77).
This indicates that the two data sources are, at the very least, not inconsistent. 
However, more fine-grained data is needed to support stronger conclusions.

The growing distance between cumulative lower and upper bands in \Cref{fig:compare-transparency-records-and-astr} from 2009 to 2013 is due to the growing number of published transparency reports.
Initially, only Google and Twitter published reports.
There was a sharp spike in the number of companies publishing reports around 2013 and 2014, presumably in response to the Snowden revelations drawing increased attention to transparency.
Some companies have stopped issuing reports since then, but the decrease in reports in 2022, which is reflected in the decreased width of the band, is likely an artifact of delayed transparency report publishing by companies.

Diving into the reported ranges of NSLs themselves, we find that 13 out of 41 reporting companies explicitly stated that they had never received any NSLs.
The vast majority of companies report the lowest band for NSL requests (i.e., 0-249, 0-499, or 0-999). 
Telecommunication companies, such as AT\&T, T-Mobile, and Verizon, received the most NSLs, with up to 1000-2000 requests each year. 
Apple and Google, both large producers of phones and mobile operating systems, were the only other companies reporting a higher range of NSL requests. 
We surmise that telecommunications and related companies may lead these statistics because they collect valuable metadata for investigations.\footnote{
    In comparison, companies like Adobe maybe be less likely to be utilized by malicious actors.
}

Due to legal restrictions regarding the discussion of NSLs by recipient companies, we were not able to reach out to any companies to ask more detailed questions about specific NSLs or general patterns regarding NSLs that they have observed. We were, however, able to investigate the subset of NSLs that have been published by some companies once the NSLs are no longer under nondisclosure restrictions.
Analyzing these NSLs is the focus of \Cref{subsec:company_nsls}.

\subsection{Company NSLs}\label{subsec:company_nsls}

The third and final data source are NSLs themselves, voluntarily published by their recipients\footnote{%
 Two avenues for publishing NSLs are the reciprocal notice procedure and the termination procedure.
} 
after the nondisclosure requirement (colloquially ``gag order'') has been lifted. 
These letters contain specific ROIs and have a gag order attached, restricting the receiving party from publishing the NSLs or even discussing their existence.
The most sensitive parts of the letters are redacted, including user identities and confidential information such as Social Security Numbers.
However, unredacted information includes the types of information requested (e.g., email metadata or credit records) and types of ROIs (i.e., user identifiers such as email addresses, account numbers, addresses, or names). 
Additionally, we found file numbers and the issuance date (which are often not redacted) to be useful metadata, in addition to the number of redacted lines that indicate the ROI volume. Finally, most companies report the publication date on which they released the redacted NSL.

Google and Apple are the largest publishers of NSL letters in our data set with 272 and 45 letters, respectively.
Appendix~\ref{subsec:data-src-nsls} lists the full composition of our data set.
We only see a fraction of all existing letters:
Between Jan 1, 2015 and Dec 31, 2020\footnote{%
    We pick these years as they are after the gag order was weakened but not too recent include currently-active gag orders.
}, companies published only 0.3\% of all issued NSLs.

We use the number of days between the issuance and publication dates as an estimator for the duration of the gag order. 
The internal administrative delay between lifting a gag order and publishing the letter may add some noise to this metric.

\begin{figure}[t]
    \centering
    \includegraphics[width=\linewidth]{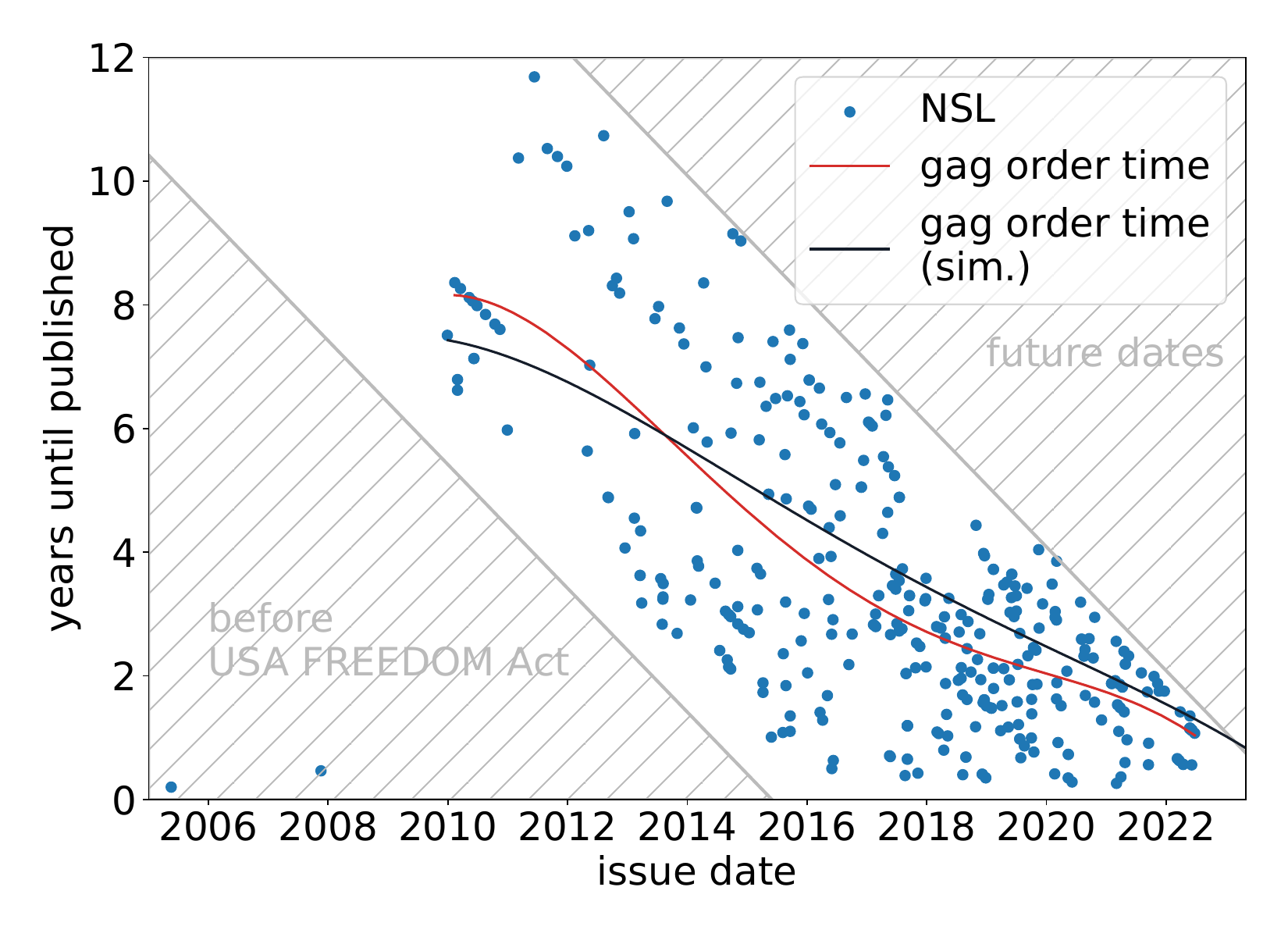}
    \caption{%
        This graph shows how many years it took to publish NSLs after they were issued.
        Most letters were only published after the USA FREEDOM Act added procedures to challenge gag orders.
        The gag order time appears to decrease.
        However, the distribution of the unpublished 99.7\% of the NSLs is unknown.
        Moreover, the measured gag order time appears to decrease due to a bias, as letters from recent years with longer times have not been published yet.
        For comparison, we plot a black line to show a trend for a simulated scenario where gag order time is sampled from the uniform distribution with the constant mean value of 5 years.}
    \label{fig:gag_order_scatter_plot}
    \Description[A graph plotting NSLs with their issuance date on the x axis and the number of years until they were published on the y axis.]{
        This graph shows how many years it took to publish NSLs after they were issued.
        Most letters were only published after the USA FREEDOM Act added procedures to challenge gag orders.
        The gag order time appears to decrease.
        However, the distribution of the unpublished 99.7\% of the NSLs is unknown.
        Moreover, the measured gag order time appears to decrease due to a bias, as letters from recent years with longer times have not been published yet.
        For comparison, we plot a black line to show a trend for a simulated scenario where gag order time is sampled from the uniform distribution with the constant mean value of 5 years.
    }
\end{figure}

The scatter plot in \Cref{fig:gag_order_scatter_plot} shows how many years after an NSL's issuance it is published.
We observe that, except for two NSLs, all letters were published after June 2015, when the USA FREEDOM Act~\cite{law:PL-114-23-FREEDOM} amended NSL reporting and nondisclosure regulations, limited gag orders, and simplified the disclosure process.
This supports the hypothesis that in the absence of public scrutiny, the gag orders before 2015 were easier to enforce and very rarely lifted.
After Congress relaxed the nondisclosure requirements, publishing NSLs became possible for companies.
It appears that few companies retroactively published NSLs that they received before these legal reforms.

There are two important caveats related to \Cref{fig:gag_order_scatter_plot}.
First, we only see 0.3\% of all NSLs, and it is unknown whether they are a representative sample for the distribution of the remaining 99.7\% of NSLs.
Second, data since 2013 is increasingly skewed towards shorter gag order times, as any letter that is in the ``future dates'' area will only be published in the future.
For instance, letters issued in 2022 with a gag order time of four years will only be published in 2026.
We characterize this bias with the simulated gag order time (black line) in \Cref{fig:gag_order_scatter_plot} compared to the polynomial regression of observed gag order times (red line). 
Furthermore, we provide a toy model to estimate the true, unbiased mean gag order duration time in \Cref{sec:gag-estimation}.




\section{Data Inconsistencies}\label{sec:inconsistencies}

This section discusses inconsistencies in the reported NSL data when comparing different sources.
Furthermore, we point out open questions that oversight bodies may consider investigating.

\subsection{Diverging Requests for Information Counts}\label{subsec:inconsistent_rois}

The Office of the Director of National Intelligence (ODNI) started to publish an Annual Statistical Transparency Report (ASTR) in 2014~\cite{astr} including the number of issued NSLs and---similar to FISA~\cite{fisa-reports}---NSL requests.
The ASTR reports were a reaction to the first leaked documents by Snowden~\cite{guardian-snowden}, ordered in June 2013~\cite{astr:2013} to increase transparency.

\begin{table*}
    \footnotesize
    \begin{center}
        \caption{%
            The number of reported Request of Information (ROI) according to different sources.
            We sum the first three columns to derive the total number of ROIs reported under FISA and compare them with the ASTR counts reported by the ODNI.
            Numbers corrected after we reported the inconsistencies are shown in parentheses.
            \label{table:roi_inconsistency}
        }
        \begin{tabular}{|c|c|c|c|c|c|} 
             \hline
                 \textbf{Year} & \textbf{US ROIs} & \textbf{Non-US ROIs} & \textbf{Subscriber ROIs} & \textbf{FISA ROIs} & \textbf{ASTR ROIs} \\
            \hline
            \hline
                2015 & 9418  & 31863  & 7361  & 48642                       & 48642 \\
            \hline
                2016 & 8727  & 6651   & 9423  & 24801  & 24801 \\
            \hline
                2017 & 9006  & 14861  & 17712 & 41579                       & 41579 \\
            \hline
                2018 & 11454 & 14481  & 12937 & 38872                       & 38872 \\
            \hline
                2019 & 8557  & 35848  & 19601 (corr.\ 19061) & \cellcolor{orange!25}64006 (corr.\ 63466)  & \cellcolor{orange!25}63466 \\
            \hline
                2020 & 6670  & 6187   & 11368 & 24225                       & 24225  \\
            \hline
                2021 & 7607  & 9486   & 14732 & \cellcolor{orange!25}31825  & \cellcolor{orange!25}39214 (corr.\ 31825) \\
            \hline
                2022 & 8587  & 9103   & 14927 & 32617                       & 32617 \\
            \hline
        \end{tabular}
        
    \end{center}
\end{table*}

The columns ``FISA ROIs'' and ``ASTR ROIs'' of \Cref{table:roi_inconsistency} show the reported number of NSL requests. 
The ``FISA ROIs'' column is the sum of the reported NSL ROIs for US persons, non-US persons, and subscriber information under FISA in the previous three columns.
Before 2015, the FISA numbers did not include non-US persons and subscriber information and can, therefore, not be compared to the ASTR numbers.

There were no public statements (or observations) of the differing numbers in 2019 and 2021, despite that the number of ASTR ROIs is over 23\% higher than the count of FISA ROIs.
After reporting these inconsistencies to the ODNI, the National Security Division issued a correction of the ROIs for subscriber information from 2019, which should have been 19,061 instead of 19,601~\cite{fisa-correction-2020}.
Moreover, we were informed that the numbers of the ASTR report for 2021 will be corrected when the ASTR report for 2024 is issued in Spring 2025. The correct number would have been 31,825, as reported by FISA.

\subsection{NSL Metadata}\label{subsec:file_nrs}

\begin{figure}[t]
    \centering
    \includegraphics[width=\linewidth]{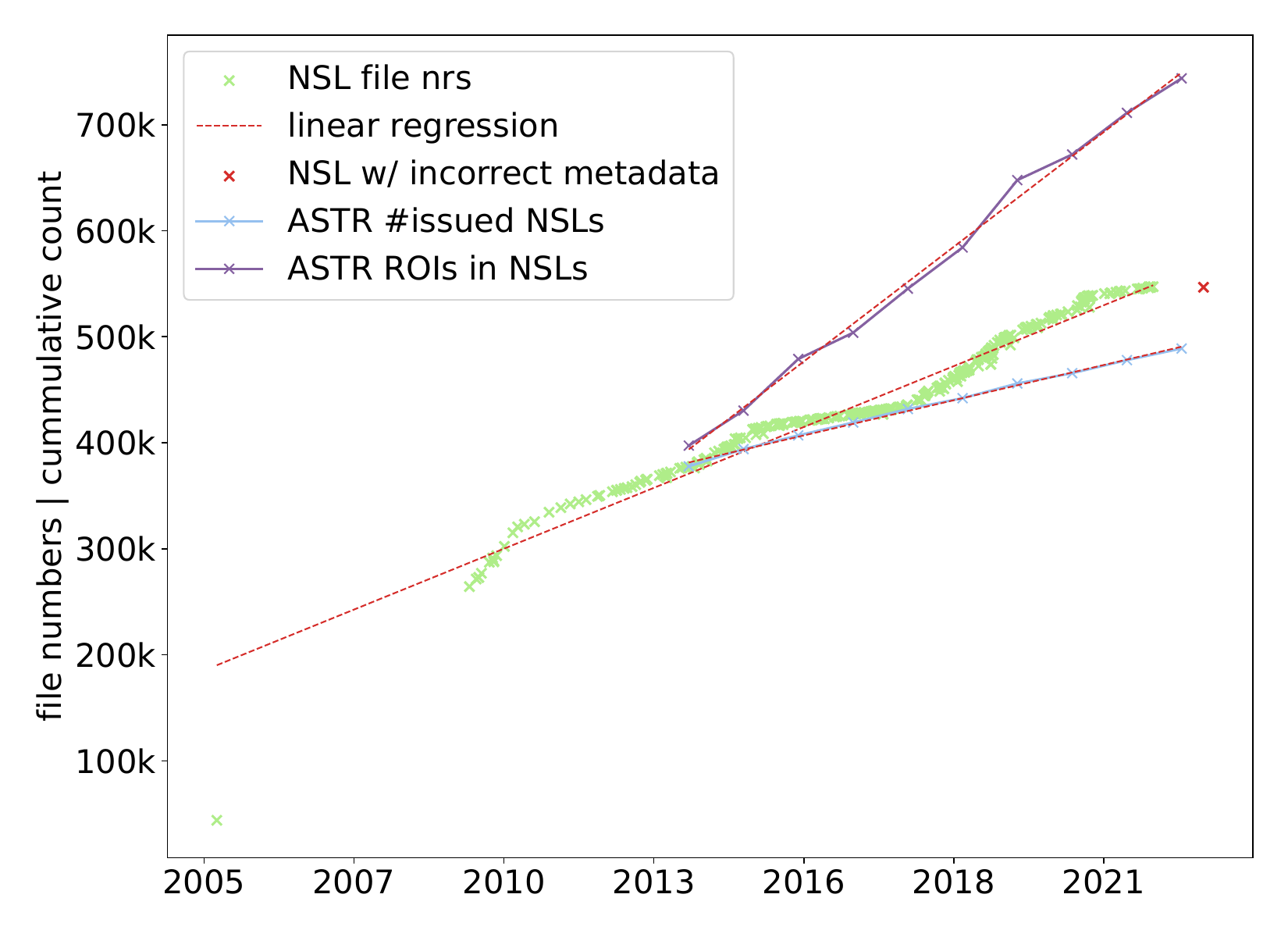}
    \caption{%
        The file numbers of NSLs plotted over time show a monotonically increasing sequence of points.
        We compare them to the cumulative ROI and NSL counts published in the ASTR, after adjusting their y-axis offset to match the number of files in 2014.
        File numbers with red crosses have erroneous issue dates reported by companies. 
    }
    \label{fig:file_nrs}
    \Description[Plot showing the file numbers respectively the cumulative count of the number of issued NSLs on the y axis and time on the x axis.]{
        The file numbers of NSLs plotted over time show a monotonically increasing sequence of points.
        We compare them to the cumulative ROI and NSL counts published in the ASTR, after adjusting their y-axis offset to match the number of files in 2014.
        Some file numbers are outliers and marked with red crosses. They had erroneous issue dates reported by companies. 
    }
\end{figure}

File numbers are part of the scarce metadata of NSL letters.
\Cref{fig:file_nrs} plots the file numbers (y-axis) over the issuance date of the letter, for all NSLs that were published by companies after their gag order was lifted.
It also plots the cumulative number of NSL requests reported by ASTR, and the number of issued NSLs.\footnote{%
    The x-axis labels mark the start of a year and we place the cumulative count at the end of a year.
    For example, the number of NSLs issued in 2015 is added to the cumulative count on Dec 31, 2015.
    Hence, it appears closer to the 2016 label. 
}
The dashed red lines are the linear regression of the respective data.
NSLs were issued since their authorization in 1978 and, therefore, the reporting that started in 2005 and 2013 respectively have an unknown offset defined by the sum of previous NSL counts of the same type.\footnote{
    Early issues with the FBI tracking system add further uncertainty.
    The OIG reports that NSLs have been issued under the case file numbers of another division~\cite{ig-report-2014}. 
}
For ease of comparison with our hypothesis, we set the y-axis offset for these lines to the last previously known file number (for instance, for data starting in 2013, the last file number of an NSL from 2012 defines the y-axis offset).

We note that the file numbers are increasing roughly monotonic where 22\% of the numbers deviate and show slight decreases of 0.4\% on average.
It seems to be a reasonable hypothesis that NSL letters are assigned consecutive numbers by the FBI before their issuance (with some being issued faster than others afterward).
The blue line of issued NSLs as reported by ASTR grows slower than the number of files.
It may be the case that some of the file numbers are assigned to NSLs that the FBI prepared to issue but withdrew before serving them to a company.

Qualitatively, we observe that there appears to be some correlation between the cumulative number of ROIs and the number of files, delayed by a few months.
The file numbers have two steeper increases, one from 2014--2015 and another from 2018--2019.
The ROI counts show similar characteristics in 2015 and 2019 (recall, the cumulative counts are reported closer to the ticks in 2016 and 2020).
These increases in the ROI counts are mainly caused by the spikes in the NSL requests for non-US persons, as \Cref{table:roi_inconsistency} shows.
However, we observe the possible anomaly that these steeper increases in file numbers are not reflected in the reported number of NSLs, despite appearing to be present in the cumulative ROI counts.
It is expected that file numbers follow trends in the number of issued NSLs and not ROIs, since a single NSL can contain multiple ROIs but should have only one file number.
One possible explanation could be a difference in the counting methodology of ROIs for non-US persons. 
The ODNI did not comment on this inconsistency.
Without internal insight into the assignment of file numbers, we are limited to observing these unexplained inconsistencies, which lowers our confidence in the accuracy of the reported data.

\section{Current Challenges and Potential Paths Forward}
In this section, we start by discussing the current operational challenges towards providing better transparency. 
We then describe potential paths forward to improve upon the current state of affairs.

\subsection{Understanding Current Challenges}
We conducted two informal interviews with the current and former Chief, ODNI Civil Liberties, Privacy, and Transparency Office to validate our findings and understand the process and challenges of government data curation.
These interviews were specific to this situation and the ODNI component that we engaged with.  We do not know the extent to which the sentiments expressed to us fully generalize to other Federal agencies in the national security sphere.  We summarize our understanding of this ODNI perspective on NSL reporting below.


An important challenge for the national security community is that publishing data about intelligence collection must be evaluated for its potential risk to security.  As documented in "Principles of Intelligence Transparency for the Intelligence Community"~\cite{odni:ic-principles} published by the ODNI,
there is a requirement to balance between transparency and security to avoid accidental disclosure of sensitive information.  Among the concerns expressed to us is metadata leakage---that a document may implicitly embed sensitive information in its digital metadata.  Thus, as seen in our data, many NSLs are published as PDF scans, specifically to exclude a potential metadata information channel.  Unfortunately, this also significantly hinders the parsing of the information that was intentionally released in the NSL report.
Another concern is the unauthorized disclosure of classified information via the content itself (reflected in the redaction in some of the PDFs).
Last but not least, publishing too much or too specific data might lead to accidental information leaks (e.g., that correlating or aggregating might allow a third-party to learn information that they should not be able to), which is a serious concern for the intelligence community.

Another practical challenge is the lack of a unified system and methodology for counting NSLs.
On the one hand, definitions of what should be counted are sometimes vaguely defined in the statute, which can lead to ambiguity in counting even within a particular organization.
On the other hand, different agencies reporting on the same data may have different methodologies for counting.
Moreover, the current system for tracking NSLs is not set up to support counting.

A third challenge explained to us was the lack of resources for such government processes. It is already a labor-intensive process to collect the numbers and minimize the risk of accidental information leaks, not to mention conducting additional analysis to ensure correctness.  
Further exacerbating this issue, it was mentioned that different agencies sometimes need to report data over different time periods, which appears to limit the ability to correlate or normalize processes.

Finally, as a meta-challenge, we as independent researchers have limited ability to evaluate the choices or tradeoffs made here because the underlying security concerns are inherently opaque and difficult to judge without access to classified information.

In summary, ODNI believes that providing better transparency and data remains a challenging task, as it can require a delicate balancing between transparency, security, resource allocation, and other operational challenges.  Moreover, identifying whether existing transparency mechanisms are achieving their policy goals and whether substantive expansion might be achievable without security impact, may require a perspective only available within the national security community.\footnote{We wonder if an organization such as the Privacy and Civil Liberties Oversight Board (PCLOB) might not offer an appropriate vehicle for such an evaluation, as its charter allows to investigate such questions in this environment.}

\subsection{Potential Paths Forward}

Given these limitations, and based on our insights from studying NSL data collection and its challenges, we make the following recommendations for the publication of future transparency data.

First, it is feasible to significantly increase the \emph{usability} of existing published data, and hence the ability of the public to interpret and review published data, without jeopardizing the security goals of the intelligence community.
We suggest publishing data in text files, as comma-separated values (the ``CSV'' file type).
This document type has no metadata, but is machine-readable and, hence, easy to process.
Intelligence.gov already maintains a public dataset, and it can be considered to expand this to other agencies.
Indeed, this is how mandated wiretap reporting is structured\footnote{\url{www.uscourts.gov/data-news/reports/statistical-reports/wiretap-reports}}---a dataset with similar concerns about inadvertent disclosure.
Second, we recommend the Congress consider consolidating data collection efforts.
Currently, the DOJ and ODNI both collect and report NSL data.
This not only duplicates counting efforts, but different reporting periods and ambiguous definitions---e.g., about scope and categorization of NSLs---may introduce discrepancies in the reported data.  A single reporting agency could better normalize reporting and reduce challenges in interpretation.  
Third, NSL data reported under FISA has little context, which is essential for interpretation.
For instance, our analysis showed that the number of requests for information fluctuate over the years.
For instance, the 39,214 ROIs reported by ASTR in 2021 may sound plausible in isolation, but knowing that the previous year only had 24,225 may be helpful to spot anomalies. (And in this specific case, could have helped to identify that the reported number was indeed several thousand ROIs too high.)
Without the historical data, longitudinal trends cannot be observed and the reported data cannot be put into perspective.
If possible, commentary on anomalies over the years would help interpreting unexpected peaks. 

We acknowledge that these changes may be challenging to achieve for existing processes. 
However, we hope our empirical analysis of the effectiveness of existing NSL transparency measures will inform future debates of the Congress on transparency regulation and reporting mandates.
These discussions can be complemented by research on alternative data reporting requirements for optimal auditability as well as the government's knowledge of what data is sensitive to national security.

\section{Discussion and Conclusion}\label{section:discussion}
Transparency is an aspirational tool for discouraging the government from abusing its power. A common way to balance transparency with secrecy is through reporting sensitive data in aggregate forms. In this way, the public can review the broad actions of the government and flag significant violations without revealing the details of individual activities. However, the success of this approach largely relies on two assumptions being true---that the data is provided in a well-documented and usable fashion, and that the public (or portions thereof) can and is regularly auditing the data to ensure compliance.

In this work, we have explored the extent to which these two assumptions hold for NSL usage in practice. By compiling a broad array of data published by both the public and private sectors, we are able to holistically evaluate the current state of data transparency of NSLs.  We further highlight that while data has been released to provide transparency to the public, the usability of the data leaves something to be desired. This lack of usability in turns hinders any public effort to audit the published data. Indeed, as part of the process, we have uncovered data discrepancies that were corrected after reporting. 

Our findings suggest that both assumptions may be too strong in practice. Indeed, our work is not the only one to make such observations. Other researchers have identified concerns about the trustworthiness and completeness of the wiretap reports long mandated by 18 U.S.C~\S~2519 (see Gidari~\cite{web:wiretap-data} and Varner and Ng~\cite{news:missing-warrants}). These concerns are echoed by the findings of an internal investigation conducted by the Federal Judicial Center, which found the government's wiretap reports to be inaccurate and incomplete~\cite{fjc:wiretap-report}. The underlying causes were multifaceted, but included a combination of procedure challenges, operational complexities, ignorance, and imperfect incentives.  Thus, while transparency may have the potential to encourage compliance and enable public oversight, this promise is clearly predicated on effective execution---which may be undermined by a range of practical and organizational factors.

\section*{Acknowledgments}\label{section:ack}
We thank our anonymous reviewers and shepherd James Grimmelmann for their valuable comments and suggestions. We thank Alisha Ukani and Stewart Grant for supporting us in interpreting our data with their data processing and visualization expertise. Additionally, we are grateful to our interviewees at ODNI for their time and sharing their insights. Finally, we thank Peter Swire and Sunoo Park for their feedback on the draft.
\bibliographystyle{plain}
\bibliography{refs}

\begin{thebibliography}{10}

\bibitem{law:PL-80-253-NSA}
{N}ational {S}ecurity {A}ct of 1947.
\newblock \url{https://govtrackus.s3.amazonaws.com/legislink/pdf/stat/61/STATUTE-61-Pg495.pdf}, July 1947.
\newblock Visited on June 12, 2023.

\bibitem{law:USC-15-1681-FCRA}
15 {U.S.\ Code} {\S} 1681 -- congressional findings and statement of purpose.
\newblock \url{https://www.law.cornell.edu/uscode/text/15/1681}, October 1970.
\newblock Visited on September 29, 2023.

\bibitem{law:USC-12-3414-RFPA-special-procedures-1978}
12 {U.S.\ Code} {\S} 3414 -- special procedures.
\newblock \url{https://www.law.cornell.edu/uscode/text/12/3414}, November 1978.
\newblock Visited on June 12, 2023.

\bibitem{law:USC-12-Chapter-35-RFPA}
12 {U.S.\ Code} {C}hapter 35 -- {R}ight to {F}inancial {P}rivacy {A}ct ({RFPA}).
\newblock \url{https://www.law.cornell.edu/uscode/text/12/chapter-35}, November 1978.
\newblock Visited on June 9, 2023.

\bibitem{law:fisaReport}
50 {U.S.\ Code} {\S} 1807 -- report of electronic surveillance.
\newblock \url{https://www.law.cornell.edu/uscode/text/50/1807}, October 1978.
\newblock Visited on June 9, 2023.

\bibitem{law:HRep-99-690}
{H}.{R}ept 99-690.
\newblock \url{https://www.congress.gov/congressional-report/99th-congress/house-report/690}.
\newblock Visited on June 12, 2023.

\bibitem{law:USC-12-3414-RFPA-special-procedures-1986}
12 {U.S.\ Code} {\S} 3414 -- special procedures.
\newblock \url{https://www.law.cornell.edu/uscode/text/12/3414}, October 1986.
\newblock Visited on June 12, 2023.

\bibitem{law:USC-18-2701-ECPA-access-stored-communications}
18 {U.S.\ Code} {\S} 2701 -- unlawful access to stored communications.
\newblock \url{https://www.law.cornell.edu/uscode/text/18/2701}, October 1986.
\newblock Visited on June 12, 2023.

\bibitem{law:USC-18-2709-ECPA-access-to-records}
18 {U.S.\ Code} {\S} 2709 -- counterintelligence access to telephone toll and transactional records.
\newblock \url{https://www.law.cornell.edu/uscode/text/18/2709}, October 1986.
\newblock Visited on June 12, 2023.

\bibitem{law:PL-99-569-intelligence-auth-act-1987}
{H.R.\ }4759 (99th): Intelligence authorization act for fiscal year 1987.
\newblock \url{https://www.govtrack.us/congress/bills/99/hr4759/text}, October 1986.
\newblock Visited on June 12, 2023.

\bibitem{law:PL-99-508-ECPA}
{H.R.\ }4952 (99th): {E}lectronic {C}ommunications {P}rivacy {A}ct of 1986.
\newblock \url{https://www.govtrack.us/congress/bills/99/hr4952/text}, October 1986.
\newblock Visited on June 12, 2023.

\bibitem{law:USC-50-3162-codified-IAA}
50 {U.S.\ Code} {S}ubchapter {VI} -- access to classified information.
\newblock \url{https://www.law.cornell.edu/uscode/text/50/chapter-44/subchapter-VI}, October 1994.
\newblock Visited on June 9, 2023.

\bibitem{law:PL-103-359-intelligence-auth-act-1995}
{H.R.\ }4299 (103rd): Intelligence authorization act for fiscal year 1995.
\newblock \url{https://www.govtrack.us/congress/bills/103/hr4299/text}, September 1994.
\newblock Visited on June 12, 2023.

\bibitem{law:PL-104-93-intelligence-auth-act-1996}
{H.R.\ }1655 (104rd): Intelligence authorization act for fiscal year 1996.
\newblock \url{https://www.govtrack.us/congress/bills/104/hr1655/text}, December 1995.
\newblock Visited on June 12, 2023.

\bibitem{law:PL-108-177-PATRIOT-section-374}
{H.R.\ }2417 (108th): Intelligence authorization act for fiscal year 2004. section 374.
\newblock \url{https://www.govtrack.us/congress/bills/108/hr2417/text}, October 2001.

\bibitem{law:PL-107-56-PATRIOT}
{U}niting and {S}trengthening {A}merica by {P}roviding {A}ppropriate {T}ools {R}equired to {I}ntercept and {O}bstruct {T}errorism ({USA PATRIOT ACT}).
\newblock \url{https://www.congress.gov/107/plaws/publ56/PLAW-107publ56.pdf}, October 2001.

\bibitem{law:PATRIOT:amendment:1510}
18 {U.S.\ Code} {\S} 1510 -- obstruction of criminal investigations.
\newblock \url{https://www.law.cornell.edu/uscode/text/18/1510}, March 2006.

\bibitem{law:PATRIOT:amendment:3511}
18 {U.S.\ Code} {\S} 3511 -- judicial review of requests for information.
\newblock \url{https://www.law.cornell.edu/uscode/text/18/3511}, March 2006.

\bibitem{law:PL-109-178-PATRIOT-2005}
S.\ 2271 (109th): {USA PATRIOT} act additional reauthorizing amendments act of 2006.
\newblock \url{https://www.govtrack.us/congress/bills/109/s2271/text}, March 2006.

\bibitem{law:PL-109-177-PATRIOT-2005}
{USA PATRIOT} improvement and reauthorization act of 2005.
\newblock \url{https://www.congress.gov/109/plaws/publ177/PLAW-109publ177.pdf}, March 2006.

\bibitem{law:PL-109-177-PATRIOT-2005-section-119}
{USA PATRIOT} improvement and reauthorization act of 2005. {Section} 119.
\newblock \url{https://www.congress.gov/109/plaws/publ177/PLAW-109publ177.pdf}, March 2006.

\bibitem{law:PL-109-177-PATRIOT-2005-section-128}
{USA PATRIOT} improvement and reauthorization act of 2005. {Section} 128.
\newblock \url{https://www.congress.gov/109/plaws/publ177/PLAW-109publ177.pdf}, March 2006.

\bibitem{dojagreement:2014}
Letter from {J}ames {M}.\ {C}ole.
\newblock \url{https://www.justice.gov/iso/opa/resources/366201412716018407143.pdf}, January 2014.
\newblock Visited on September 29, 2023.

\bibitem{astr:2013}
Statistical transparency report regarding use of national security authorities.
\newblock \url{https://www.intelligence.gov/assets/documents/702\%20Documents/statistical-transparency-report/National_Security_Authorities_Transparency_Report_CY2013.pdf}, June 2014.

\bibitem{law:HRep-114-109}
{H}.{R}ept 114-109 (2015).
\newblock \url{https://www.govinfo.gov/content/pkg/CRPT-114hrpt109/pdf/CRPT-114hrpt109-pt1.pdf}.
\newblock Visited on June 12, 2023.

\bibitem{law:USC-50-1874-NSL-ranges}
50 {U.S.\ Code} {\S} 1874 -- public reporting by persons subject to orders.
\newblock \url{https://www.law.cornell.edu/uscode/text/50/1874}, June 2015.
\newblock Visited on June 15, 2023.

\bibitem{law:PL-114-23-FREEDOM}
{U}niting and {S}trengthening {A}merica by {F}ulfilling {R}ights and {E}nsuring {E}ffective {D}iscipline over {M}onitoring ({FREEDOM ACT}) act of 2015.
\newblock \url{https://www.congress.gov/bill/114th-congress/house-bill/2048/text}, June 2015.
\newblock Visited on June 9, 2023.

\bibitem{wikipedia:patriot-sept-11}
{USA PATRIOT} act.
\newblock \url{https://en.wikipedia.org/wiki/Patriot_Act}.
\newblock Visited on June 12, 2023.

\bibitem{beller2022401}
Sarah Beller.
\newblock 401--forbidden: An empirical study of {Foreign Intelligence Surveillance Act} notices, 1990--2020.
\newblock {\em Harvard National Security Journal}, 13:158--223, 2022.

\bibitem{bloch2016process}
Hannah Bloch-Wehba.
\newblock Process without procedure: {N}ational {S}ecurity {L}etters and {F}irst {A}mendment rights.
\newblock {\em Suffolk University Law Review}, 49:367--408, 2016.

\bibitem{bloch2018exposing}
Hannah Bloch-Wehba.
\newblock Exposing secret searches: A {First Amendment} right of access to electronic surveillance orders.
\newblock {\em Washington Law Review}, 93:145--200, 2018.

\bibitem{fisa-correction-2020}
Slade Bond.
\newblock Correction to the {July} 17, 2020 report.
\newblock \url{https://www.justice.gov/nsd/media/1355736/dl?inline}.
\newblock Visited on September 30, 2024.

\bibitem{dallal2018speak}
Rachel Dallal.
\newblock Speak no evil: {N}ational {S}ecurity {L}etters, gag orders, and the {F}irst {A}mendment.
\newblock {\em Berkeley Technology Law Journal}, 33:1115--1146, 2018.

\bibitem{CRS:legalBackground}
Charles Doyle.
\newblock {N}ational {S}ecurity {L}etters in foreign intelligence investigations: Legal background.
\newblock \url{https://repository.library.georgetown.edu/bitstream/handle/10822/1053734/RL33320.pdf?sequence=1}, July 2015.
\newblock RL33320 Version 18.

\bibitem{EPICNati14:online}
{Electronic Privacy Information Center}.
\newblock {EPIC} -- {National Security Letters}.
\newblock \url{https://archive.epic.org/privacy/nsl/}, 2025 01.
\newblock Visited on January 18, 2025.

\bibitem{frankle2018practical}
Jonathan Frankle, Sunoo Park, Daniel Shaar, Shafi Goldwasser, and Daniel Weitzner.
\newblock Practical accountability of secret processes.
\newblock In {\em 27th USENIX Security Symposium (USENIX Security 18)}, pages 657--674, 2018.

\bibitem{garlinger2009privacy}
Patrick~P Garlinger.
\newblock Privacy, free speech, and the {PATRIOT} act: {F}irst and {F}ourth {A}mendment limits on {N}ational {S}ecurity {L}etters.
\newblock {\em NYUL Rev.}, 84:1105, 2009.

\bibitem{web:wiretap-data}
Albert Gidari.
\newblock Wiretap reports not so transparent.
\newblock \url{https://cyberlaw.stanford.edu/blog/2017/01/wiretap-reports-not-so-transparent/}, January 2017.

\bibitem{gorham2008national}
Ursula Gorham-Oscilowski and Paul~T Jaeger.
\newblock {N}ational {S}ecurity {L}etters, the {USA PATRIOT} act, and the {C}onstitution: The tensions between national security and civil rights.
\newblock {\em Government Information Quarterly}, 25(4):625--644, 2008.

\bibitem{guardian-snowden}
Glenn Greenwald.
\newblock {NSA} collecting phone records of millions of {Verizon} customers daily.
\newblock {\em The Guardian}, June 2013.
\newblock \url{https://www.theguardian.com/world/2013/jun/06/nsa-phone-records-verizon-court-order}.

\bibitem{kesari2023data}
Aniket Kesari.
\newblock Do data breach notification laws work?
\newblock {\em New York University Journal of Legislation and Public Policy}, 26:173--238, 2023.

\bibitem{odni:ic-principles}
Jason Klitenic.
\newblock Intelligence community legal reference book.
\newblock \url{https://www.dni.gov/files/documents/OGC/IC%20Legal%20Reference%20Book%202020.pdf}.
\newblock Winter 2020.

\bibitem{krollsecure}
Joshua~A Kroll, Edward~W Felten, and Dan Boneh.
\newblock Secure protocols for accountable warrant execution.
\newblock \url{http://www.cs.princeton.edu/felten/warrant-paper.pdf/}, 2014.

\bibitem{case:DoeVMukasey}
Jon~O. Newman.
\newblock 549 {F.3d} 861 ({Second Circuit} 2008). {D}oe v.\ {M}ukasey (formerly known as {D}oe v.\ {A}shcroft).
\newblock \url{https://www.aclu.org/sites/default/files/pdfs/safefree/doevmukasey_decision.pdf}, August 2008.

\bibitem{nieland2006national}
Andrew~E Nieland.
\newblock {N}ational {S}ecurity {L}etters and the amended {PATRIOT} act.
\newblock {\em Cornell L. Rev.}, 92:1201, 2006.

\bibitem{astr}
Office of~Civil Liberties{,}~Privacy{,} and Transparency.
\newblock Annual statistical transparency report.
\newblock \url{https://www.intel.gov/annual-statistical-transparency-report}.
\newblock Visited on June 9, 2023.

\bibitem{fisa-reports}
U.S.~Department of~Justice.
\newblock {FISA} reports.
\newblock \url{https://www.justice.gov/nsd/fisa-reports}.
\newblock Visited on June 9, 2023.

\bibitem{ig-report-2006}
Office of~the Inspector~General.
\newblock A review of the {F}ederal {B}ureau of {I}nvestigation's use of {N}ational {S}ecurity {L}etters in 2006.
\newblock \url{https://oig.justice.gov/sites/default/files/legacy/special/s0803b/final.pdf}, March 2008.

\bibitem{ig-report-2014}
Office of~the Inspector~General.
\newblock A review of the {F}ederal {B}ureau of {I}nvestigation's use of {N}ational {S}ecurity {L}etters.
\newblock \url{https://oig.justice.gov/reports/2014/s1408.pdf}, August 2014.

\bibitem{ig-report-2007}
Office of~the Inspector~General.
\newblock A review of the {F}ederal {B}ureau of {I}nvestigation's use of {N}ational {S}ecurity {L}etters, {M}arch 2007 (january 8, 2016 version).
\newblock \url{https://www.oversight.gov/sites/default/files/oig-reports/o1601b.pdf}, January 2016.

\bibitem{business-records-ig-review}
Office of~the Inspector General. U.S.\ Department~of Justice.
\newblock A review of the {FBI}'s use of {S}ection 215 orders for business records in 2012 through 2014.
\newblock \url{https://oig.justice.gov/reports/2016/o1604.pdf}, September 2016.
\newblock Visited on June 10, 2023.

\bibitem{fjc:wiretap-report}
David Rauma, James Eaglin, Carly Giffin, and Marvin Astrada.
\newblock Study of the administrative office of the {U.S.} courts’ wiretap report.
\newblock {\em Federal Judicial Center}, November 2021.
\newblock \url{https://www.documentcloud.org/documents/21151429-wiretap-study-report-final}.

\bibitem{romanosky2014empirical}
Sasha Romanosky, David Hoffman, and Alessandro Acquisti.
\newblock Empirical analysis of data breach litigation.
\newblock {\em Journal of Empirical Legal Studies}, 11(1):74--104, 2014.

\bibitem{schrop2017your}
Kevin~J Schrop.
\newblock Your cooperation is greatly appreciated: The {F}ourth {A}mendment, {N}ational {S}ecurity {L}etters, and public-private data sharing.
\newblock {\em Penn St. L. Rev.}, 122:849, 2017.

\bibitem{segal2014catching}
Aaron Segal, Bryan Ford, and Joan Feigenbaum.
\newblock Catching bandits and only bandits: Privacy-preserving intersection warrants for lawful surveillance.
\newblock In {\em 4th USENIX Workshop on Free and Open Communications on the Internet (FOCI 14)}, 2014.

\bibitem{smith2009kudzu}
Stephen~Wm Smith.
\newblock {Kudzu} in the courthouse: Judgments made in the shade.
\newblock {\em Federal Courts Law Review}, 3:177--216, 2009.

\bibitem{smith2012gagged}
Stephen~Wm Smith.
\newblock Gagged, sealed \& delivered: Reforming {ECPA}'s secret docket.
\newblock {\em Harvard Law \& Policy Review}, 6:313--338, 2012.

\bibitem{case:DoeVGonzales}
D.~Connecticut United States District~Court.
\newblock 386 {F.Supp.2d} 66 (2005). {D}oe v.\ {G}onzales.
\newblock \url{https://www.leagle.com/decision/2005452386fsupp2d661441}, September 2005.

\bibitem{case:DoeVAshford}
S.D. New~York United States District~Court.
\newblock 334 {F.Supp.2d} 471 (2004). {D}oe v.\ {A}shcroft.
\newblock \url{https://www.leagle.com/decision/2004805334fsupp2d4711766}, September 2004.

\bibitem{case:DoeVAshford:476}
S.D. New~York United States District~Court.
\newblock 334 {F.Supp.2d} at 476. {D}oe v.\ {A}shcroft.
\newblock \url{https://www.leagle.com/decision/2004805334fsupp2d4711766}, September 2004.

\bibitem{case:DoeVAshford:506}
S.D. New~York United States District~Court.
\newblock 334 {F.Supp.2d} at 506. {D}oe v.\ {A}shcroft.
\newblock \url{https://www.leagle.com/decision/2004805334fsupp2d4711766}, September 2004.

\bibitem{news:missing-warrants}
Maddy Varner and Alfred Ng.
\newblock Thousands of geofence warrants appear to be missing from a california {DOJ} transparency database.
\newblock {\em The Markup}, November 2021.
\newblock \url{https://themarkup.org/privacy/2021/11/03/thousands-of-geofence-warrants-appear-to-be-missing-from-a-california-doj-transparency-database}.

\bibitem{fisa-report-2008}
Ronald Weich.
\newblock 2008 annual report regarding {FISA} and the {USA PATRIOT} act.
\newblock \url{https://www.justice.gov/sites/default/files/nsd/legacy/2014/07/23/2008fisa-ltr.pdf}, May 2009.
\newblock Visited on June 9, 2023.

\bibitem{counterintelligence-2019}
Christopher Wray.
\newblock The {FBI} and the national security threat landscape: The next paradigm shift.
\newblock \url{https://www.fbi.gov/news/speeches/the-fbi-and-the-national-security-threat-landscape-the-next-paradigm-shift}, April 2019.

\end{thebibliography}

\clearpage
\appendix

\section{Additional Background for NSLs}\label{sec:full_background}
In this section, we provide additional background information for NSLs.
\subsection{Statutory Provisions for NSLs}\label{subsec:nsl-provisions}
At the time of writing, five statutory provisions grant government agencies legal authority to issue NSLs. 
Namely, the Right to Financial Privacy Act (RFPA)~\cite{law:USC-12-Chapter-35-RFPA}, (Title II of) the Electronic Communications Privacy Act (ECPA)~\cite{law:USC-18-2701-ECPA-access-stored-communications}, the National Security Act (NSA)~\cite{law:PL-80-253-NSA}, the Fair Credit Reporting Act (FCRA)~\cite{law:USC-12-Chapter-35-RFPA}, and the USA PATRIOT Act~\cite{law:PL-107-56-PATRIOT}.
We introduce each statute in chronological order.

\subsubsection{The Right to Financial Privacy Act of 1978}\label{subsec:rfpa-1978}
The RFPA, codified at 12~U.S.C.~\S~3401 \textit{et seq.}~\cite{law:USC-12-Chapter-35-RFPA}, is the first statutory provision for NSLs. 
It defines a special procedure (codified at 12~U.S.C.~\S~3414~\cite{law:USC-12-3414-RFPA-special-procedures-1978}) for government agencies to request the production and disclosure of financial records for the purposes of conducting intelligence activities. 
However, this legislation did not mandate access to such records. 
In practice, financial institutions declined requests from the FBI due to conflicts with State privacy or banking laws that prohibited the production of such records~\cite{law:HRep-99-690,CRS:legalBackground}. 
This led to the amendment of the RFPA in 1986.

\subsubsection{The Amendment to RFPA in 1986}\label{subsubsec:rfpa-1986}
In October 1986, as part of the Intelligence Authorization Act for Fiscal Year 1987~\cite{law:PL-99-569-intelligence-auth-act-1987}, Congress passed an act that amended RFPA to explicitly grant the FBI access to financial records provisioned in 12~U.S.C.~\S~3414~\cite{law:USC-12-3414-RFPA-special-procedures-1986}.
This amendment also introduced a mandate that required a semiannual report on the number of requests made to certain congressional bodies (the select committee of the House and the Senate) and a nonclosure requirement without a specific end date~\cite{law:PL-99-569-intelligence-auth-act-1987}.

\subsubsection{The Electronic Communications Privacy Act of 1986}\label{subsec:ecpa-1986}
In parallel with the amendment to RFPA, Congress passed the second statutory provision for NSLs as part of the Electronic Communications Privacy Act (ECPA) of 1986~\cite{law:PL-99-508-ECPA}. 
The Title II of the ECPA, which is called the Stored Communications Act (SCA) and codified at 18 U.S.C.~\S~2701~\textit{et seq.}~\cite{law:USC-18-2701-ECPA-access-stored-communications}, gave the FBI access to business records of wire or electronic communication service providers for counterintelligence purposes (18~U.S.C.~\S~2709~\cite{law:USC-18-2709-ECPA-access-to-records}).
Furthermore, this act defined similar non-disclosure and reporting requirements to those in \Cref{subsubsec:rfpa-1986}.

\subsubsection{The Amendment to NSA in 1994}\label{subsec:nsa-1994}
First introduced in 1947, the National Security Act~\cite{law:PL-80-253-NSA} was amended in 1994 to include the third statutory provision for NSLs (as part of the Intelligence Authorization Act for Fiscal Year 1995~\cite{law:PL-103-359-intelligence-auth-act-1995}).
This amendment introduced a procedure (codified at 50 U.S.C~\S~3162~\cite{law:USC-50-3162-codified-IAA}) for any authorized agency to request a broad array of business records from various organizations for investigating potential document leaks from government employees.
Compared to the previous two statutory provisions, this one permitted more agencies to request a wider range of documents albeit only under specific circumstances.
Additionally, unlike the first two NSL statues, it did not have a reporting mandate.

\subsubsection{The Amendment to FCRA in 1996}\label{subsec:fcra-1996}
Initially enacted in 1970, the Fair Credit Reporting Act (FCRA)~\cite{law:USC-12-Chapter-35-RFPA} was amended in 1996 to incorporate the fourth statutory provision for NSLs (as part of the Intelligence Authorization Act for Fiscal Year 1996~\cite{law:PL-104-93-intelligence-auth-act-1996}).
This amendment included a procedure (codified at 15 U.S.C.~\S~1681u~\cite{law:USC-15-1681-FCRA}) for the FBI to access credit agency records. 
Like those defined in RFPA and ECPA, it came with a reporting mandate for certain congressional bodies on a semiannual basis and allowed the FBI to demand nondisclosure.

\subsubsection{The PATRIOT Act of 2001}\label{subsec:patriot-2001}
The PATRIOT Act of 2001~\cite{law:PL-107-56-PATRIOT} was enacted as a response to the terrorist attacks on September 11~\cite{wikipedia:patriot-sept-11}.
It made substantial amendments to three of the four existing NSL statutes (RFPA, ECPA, and FCRA) and added a fifth one.
The amendments in this act extended the scope of NSLs to allow the FBI to request information for people who are not themselves a foreign power or its agent but are relevant to an investigation concerning international terrorism or foreign spying.
In addition to making NSLs more widely applicable, the amendments also simplified the approval process by authorizing Special Agents in Charge to issue NSLs in addition to officials in the FBI headquarters~\cite{CRS:legalBackground}.

The fifth NSL statute was introduced by amending FCRA. It introduced a procedure (codified at 15~U.S.C.~\S~1681v~\cite{law:USC-15-1681-FCRA}) for government agencies to access consumer reports from credit reporting agencies to conduct the investigations of intelligence or terrorist activities.
Compared to other statutes, this fifth one came with a nondisclosure requirement but not a reporting mandate (such an obligation was later added as part of a subsequent amendment).

\subsection{Subsequent Amendments}
While the five NSL statutes continue to get amended in subsequent statutes, no new NSL statute was introduced after 2001.
We briefly survey important amendments to the five existing NSL statutes after 2001.

\subsubsection{RFPA amendment in 2003}
In 2003, as part of the Intelligence Authorization Act for Fiscal Year 2004~\cite{law:PL-108-177-PATRIOT-section-374}, Congress amended the definition of financial institutions in RFPA to include a much broader range of organizations such as insurance companies and travel agencies.

\subsubsection{PATRIOT Act amendments} 
In 2006,  the 109\textsuperscript{th} Congress amended the USA PATRIOT Act with two statues, in part as a response to earlier judicial reactions to the USA PATRIOT Act~\cite{nieland2006national}.
Specifically,  two court cases \DoeVGonzales~\cite{case:DoeVGonzales} and \DoeVAshcroft~\cite{case:DoeVAshford}\footnote{
  The plaintiff's name ``John Doe'' is used in both cases because the gag order prohibited the disclosure of the plaintiff's identity.   
} raised First and Fourth Amendment issues with NSLs connected to the non-disclosure provisions and the absence of judicial oversight~\cite{CRS:legalBackground}. 
In the case of \DoeVAshcroft, the district court held that NSLs violated the Fourth Amendment because they authorized ``coercive searches effectively immune from any judicial process''~\cite{case:DoeVAshford:506}.
Moreover, the court held that the nondisclosure provisions unconstitutionally restricted free speech as they can prohibited disclosure without providing judicial means to challenge a ban or achieve eventual relief~\cite{case:DoeVAshford:506,case:DoeVAshford:476}.

As a reaction to both cases, and while \DoeVAshcroft~was on appeal at the Second Circuit, the 109\textsuperscript{th} Congress amended the USA PATRIOT Act with the USA PATRIOT Improvement and Reauthorization Act~\cite{law:PL-109-177-PATRIOT-2005} and the USA PATRIOT Act Reauthorization Amendments Act~\cite{law:PL-109-178-PATRIOT-2005}. Collectively, besides adding penalties for non-compliance with NSLs or their nondisclosure requirements~\cite{law:PATRIOT:amendment:3511,law:PATRIOT:amendment:1510}, these two statutes introduced judicial review on the use of NSLs and weakened the nondisclosure requirements.
Specifically, the amendments require that the Foreign Intelligence Surveillance Court (FISC) approves NSL requests and adds control mechanisms and a process to ease nondisclosure provisions~\cite{law:PATRIOT:amendment:3511,law:PATRIOT:amendment:1510}.
Furthermore, the changes mandated classified semiannual reports about the use of NSLs to Congress and nonclassified annual statistics reports~\cite{law:PL-109-177-PATRIOT-2005-section-128}.
The latter reports are one of the main data sources for this paper.
Regarding \DoeVAshcroft, the Second Circuit ruled in 2008 that the FBI needs to certify that disclosure of an NSL would lead to statutorily enumerated harms to justify non-disclosure provisions~\cite{case:DoeVMukasey}. Last, the amendments call for an audit on use of the NSLs from the Office of the Inspector General (referred to as the OGI reports).

\subsubsection{Office of the Inspector General (OIG) reports} 
As part of the PATRIOT Act amendments (P.L.~109-177~\S~119~\cite{law:PL-109-177-PATRIOT-2005-section-119}), the Office of Inspector General (OIG) of the Department of Justice was required to audit and report on the use of NSLs. 
On March 9, 2007~\cite{ig-report-2007}, soon after the FBI started reporting data on NSLs, the OIG released its first report. 
Notably, the report suggested that the internal processes by which the FBI tracks NSLs were flawed. 
For instance, NSLs that are issued under different statutes are tracked independently which might lead to counting the same person multiple times. 
Similar issues were mentioned in follow-up reports, noting that the same person might be counted as multiple people due to inconsistent spelling of their name. 
In response to the OIG reports, the FBI took steps to correct the identified deficiencies and introduced a new internal database for tracking NSLs. 
While they tried to retroactively correct information about NSLs for 2006 and 2007 manually, they stated that this information might still be inaccurate~\cite{fisa-report-2008}.

\subsubsection{USA FREEDOM Act}
In 2015, Congress enacted the USA FREEDOM Act~\cite{law:PL-114-23-FREEDOM}, which revised the nondisclosure requirement, introduced an additional reporting requirement, and expressly limited the use of NSL to specifically identified information. 
Notably, the additional reporting requirement and restrictions on the use of NSLs are direct responses to NSA's bulk metadata collection practice~\cite{business-records-ig-review}.

\section{Data Sources}\label{sec:data-srcs}
In this section, we list the companies for which we can find transparency reports as well as companies that publish individual NSLs.
\subsection{List of Transparency Report Companies}\label{subsec:data-src-transparency-reports}

We used transparency reports from the following companies to collect our data:
\begin{multicols}{2}
\begin{itemize}
\item 23andMe
    \item Adobe
    \item Airbnb
    \item Amazon
    \item Apple
    \item AT\&T
    \item Cisco
    \item Cloudflare
    \item Coinbase
    \item Comcast
    \item cPanel
    \item Credo
    \item Discord
    \item DreamHost
    \item Dropbox
    \item eBay
    \item Etsy
    \item Evernote
    \item Facebook
    \item GitHub
    \item Google
    \item IBM
    \item Kickstarter
    \item Lantern
    \item Let's Encrypt
    \item LinkedIn
    \item Lookout
    \item Lyft
    \item Mapbox
    \item Medium
    \item Microsoft
    \item nest
    \item Netflix
    \item Pinterest
    \item Reddit
    \item Ring
    \item Slack
    \item Snapchat
    \item Sonic
    \item Sonos
    \item SpiderOak
    \item T-Mobile US
    \item TikTok
    \item Tumblr
    \item Twilio
    \item Twitch
    \item Twitter
    \item Uber
    \item Verizon
    \item Virtru
    \item Wickr
    \item Wikimedia
    \item Word Press
    \item Yahoo
    \item Zoom
\end{itemize}
\end{multicols}

\subsection{Number of NSLs per Company}\label{subsec:data-src-nsls}

\Cref{table:nsl_dataset_stats} shows the number of National Security Letters (published after their gag order was lifted) contained in our data set.

\begin{table}
    \begin{center}
        \caption{Statistics on NSLs published by companies\label{table:nsl_dataset_stats}}
        \begin{tabular}{|c|c|} 
             \hline
                 \textbf{Company} & \textbf{\# of published NSLs}\\
            \hline
            \hline
            Google & 272 \\
            \hline
            Apple & 45 \\
            \hline
            Facebook (now Meta) & 15 \\
            \hline
            Twitter & 11 \\
            \hline
            Automattic/WordPress & 5 \\
            \hline
            Yahoo & 3 \\
            \hline
            CREDO Mobile & 2 \\
            \hline
            Internet Archive & 2 (redacted file numbers) \\
            \hline
            Twilio & 2 \\
            \hline
            Cloudflare & 1 \\
            \hline
            Library Connection & 1 \\
            \hline
            Microsoft & 1 \\
            \hline
        \end{tabular}
    \end{center}
\end{table}

\newtheorem{assumption}{Assumption}

\section{Nondisclosure duration time}\label{sec:gag-estimation}
\subsection{Bias characterization and numerical experiments}
In this section, we characterize a bias in a measured mean value of the nondisclosure duration time and suggest a simple model to estimate, under some assumptions, the true mean value.
Most importantly, we show that the decreasing measured mean value, counter intuitively, does not imply that the true mean is decreasing.

\begin{figure}[t]
    \centering
    \includegraphics[scale=0.4]{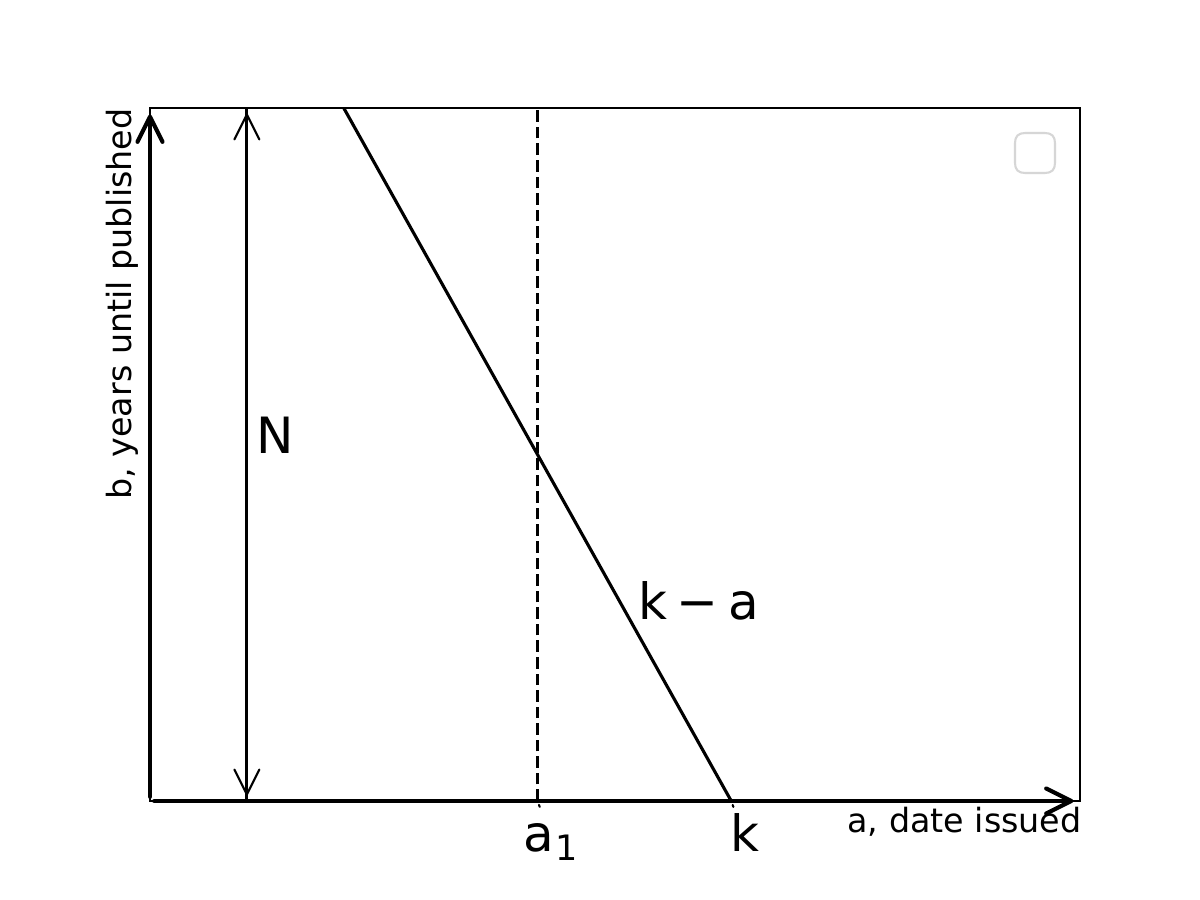}
    \caption{This graph shows schematically how we interpret the NSL letters as observations drawn from some distribution.}
    \label{fig:gibbs_graph}
    \Description[This graph shows schematically how we interpret the NSL letters as observations drawn from some distribution.]()
\end{figure}

We can interpret the data points $(a, b)$ on Figure \ref{fig:gag_order_scatter_plot} as an observations from some distribution $P_{observed}(a, b)$, where $a$ is a date of issue, $b$ is a nondisclosure duration of the NSL letter, $P_{observed}$ is the probability of a letter to become publicly available, $N$ is the maximum gag order duration time, and $k$ is a current time (see Figure \ref{fig:gibbs_graph}).

Let us denote as $P_{lifted}(b=b_1 | a=a_1)$ a conditional probability of a gag order to be lifted after time $b_1$ given the NSL is issued at time $a_1$.
To account for the selective publishing of NSL by companies, we introduce $P_{published}({published} | a=a_1, b=b_1)$, a conditional probability of a letter to be published given it is issued at time $a_1$ and the gag order is lifted at time $b_1$.
Finally, we introduce the indicator function $\mathds{1}\{a_1 + b_1 \leq k\}$ to account for the fact the we only observe NSL letters published before the current time and do not observe the future letters.
Then, $P_{observed}(a, b)$ is factorized as:

\begin{equation}
\begin{split}
P_{observed}(a, b) & \propto \mathds{1}\{a_1 + b_1 \leq k\} \\
                   & \cdot P_{published}({published} | a=a_1, b=b_1) \\
                   & \cdot P_{lifted}(b=b_1 | a=a_1).
\end{split}
\end{equation}

    

The bias in the measurement and the decreasing trend on the Figure \ref{fig:gag_order_scatter_plot} stems from the indicator function $\mathds{1}\{a_1 + b_1 \leq k\}$.
\newline
\newline

To simplify our model, we introduce 2 assumptions:

\begin{assumption}\label{as:gag_assumption_1} Probability of a letter to be published is constant:
\[
    P_{published}({published} | a=a_1, b=b_1) = p_{pub}.
\]
\end{assumption}

\begin{assumption}\label{as:gag_assumption_2} Gag order duration time is uniformly distributed:
\[
    P_{lifted}(b=b_1 | a=a_1) \sim U[0, N].
\]
\end{assumption}

Now, having defined this distribution, we can sample from it using a Gibbs sampling approach (i.e., first, sample $a$, then separately sample $b$ using Algorithm \ref{alg:cap})

\begin{algorithm}
\caption{NSL sampler}\label{alg:cap}
\begin{algorithmic}
\State $a_1 \sim U[0, k]$
\State $b_1 \sim U[0, N]$
\State $pub \sim Bernoulli(p_{pub})$
\If{$a_1 + b_1 < k$ and $pub$}
    \State return $(a_1, b_1)$
\Else
    \State return $\emptyset$
\EndIf
\end{algorithmic}
\end{algorithm}

Now, we sample observations using Algorithm \ref{alg:cap} and plot polynomial regression of the observations to obtain the simulated, black curve on Figure \ref{fig:gag_order_scatter_plot}.

\subsection{True mean estimation}

With this model, we can also account for the bias and estimate the true mean gag order duration time in the following way.

Let us fix a certain value of $a=a_1$.
In this case, the probability that sampling one NSL returns a non-empty value, i.e., that we observe the sampled NSL in our current time, is $P_{observed}$ marginalized over $b$:
\begin{equation}
\begin{split}
P_{observed}(a=a_1) & = \sum_{b \in [0, N]} P_{observed}(a=a_1, b=b_1) \\
                    & = \frac{k - a_1}{N} \cdot p_{pub}.
\end{split}
\end{equation}
Given that we have $p_{pub}$ and $k$ being constant, and $a_1$ being fixed, this marginalized $P_{observed}$ becomes a Bernoulli distribution:
\[
    P_{observed}(a=a_1) = Bernoulli(\frac{k - a_1}{N} \cdot p_{pub}).
\]
And the number of observed NSLs is therefore a Binomial distribution $B(n, \frac{k - a_1}{N} \cdot p_{pub})$
From there, we can compute its expectation by plugging $n_{a_1}$, the number of published NSLs at a time $a_1$ (the number of Bernoulli trials), which we can obtain from ASTR data:
\[
    E[B(n_{a_1}, \frac{k - a_1}{N} \cdot p_{pub})] = \frac{k - a_1}{N} \cdot p_{pub} \cdot n_{a_1}.
\]
Finally, we know empirical value of $E[B(n_{a_1}, \frac{k - a_1}{N} \cdot p_{pub})]$ from NSL data and from Figure \ref{fig:gag_order_scatter_plot}, which is the number of publicly available NSL letters issued at time $a_1$. Let us denote it as $\#obs_{a_1}$.
That gives us an expression of $N$ for certain $a=a_1$:
\[
N_{a_1} = \frac{k - a_1}{\#obs_{a_1}} \cdot p_{pub} \cdot n_{a_1}.
\]
In our tests, unfortunately, we found that this estimator does not produce plausible values of $N$, and one of the reasons might be that the actual distribution of gag order duration time $b$ is not uniform, violating our assumption \ref{as:gag_assumption_2}.
It is not impossible that an improved approach can provide robust mean estimation for this scenario.
We hope that this toy model might be a step towards understanding what information public can infer from the National Security Letters.

\end{document}